\documentclass[useAMS,usenatbib]{mn2e}
\usepackage{multirow}
\usepackage{graphicx}
\usepackage{amssymb} 
\def\aap{A\&A}

\def\apj{ApJ}
\def\apjs{ApJS}

\def\mnras{MNRAS}
\def\aj{AJ}
\def\nat{Nature}

\def\prd{Phys. Rev. D}

\def\araa{ARA\&A}       
\def\jcap{J. Cosmol. Astropart. Phys.}

\def\om{\Omega_{\rm m}}

\def \mpc {\,h^{-1}{\rm Mpc}}
\begin{document}

\title[Cosmological implications of the BOSS DR11 $\xi_{\perp}(s)$ and $\xi_{\parallel}(s)$]
{
The clustering of galaxies in the SDSS-III Baryon Oscillation
 Spectroscopic Survey: cosmological implications of the full shape of the clustering wedges in the data
release 10 and 11 galaxy samples
}
\author[A.G. S\'anchez et al.]
{\parbox[t]{\textwidth}{
Ariel~G. S\'anchez$^{1}$\thanks{E-mail: arielsan@mpe.mpg.de},
Francesco Montesano$^{1}$,
Eyal~A. Kazin$^{2,3}$,
Eric Aubourg$^{4}$,
Florian Beutler$^{5}$,
Jon Brinkmann$^{6}$,
Joel R. Brownstein$^{7}$,
Antonio~J. Cuesta$^{8,9}$,
Kyle~S. Dawson$^{7}$,
Daniel~J. Eisenstein$^{10}$,
Shirley Ho$^{5,11}$,
Klaus Honscheid$^{12}$,
Marc Manera$^{13,14}$, 
Claudia Maraston$^{14}$,
Cameron~K. McBride$^{10}$,
Will~J. Percival$^{14}$,
Ashley~J. Ross$^{14}$,
Lado Samushia$^{14,15}$,
David~J. Schlegel$^{5}$,
Donald~P. Schneider$^{16,17}$,
Ramin Skibba$^{18}$,
Daniel Thomas$^{14}$,
Jeremy~L. Tinker$^{19}$,
Rita Tojeiro$^{14}$,
David~A. Wake$^{20,21}$
Benjamin~A. Weaver$^{19}$,
Martin White$^{22,5}$
and Idit Zehavi$^{23}$
}
\vspace*{6pt} \\ 
$^{1}$ Max-Planck-Institut f\"ur extraterrestrische Physik, Postfach 1312, Giessenbachstr., 85741 Garching, Germany\\ 
$^{2}$ Centre for Astrophysics and Supercomputing, Swinburne University of Technology, P.O. Box 218, Hawthorn, Victoria 3122, Australia\\
$^{3}$ ARC Centre of Excellence for All-sky Astrophysics (CAASTRO)\\
$^{4}$ APC, University of Paris Diderot, CNRS/IN2P3, CEA/IRFU, Observatoire de Paris, Sorbonne Paris Cit\'e, France\\  
$^{5}$ Lawrence Berkeley National Laboratory, 1 Cyclotron Rd, Berkeley, CA 94720, USA\\
$^{6}$ Apache Point Observatory, P.O. Box 59, Sunspot, NM 88349-0059, USA.\\
$^{7}$ Department of Physics and Astronomy, University of Utah, 115 S 1400 E, Salt Lake City, UT 84112, USA\\
$^{8}$ Institut de Ci\`encies del Cosmos, Universitat de Barcelona, IEEC-UB, Mart\'{i} Franqu\`es 1, E08028 Barcelona, Spain \\
$^{9}$ Department of Physics, Yale University, 260 Whitney Ave, New Haven, CT 06520, USA\\
$^{10}$ Harvard-Smithsonian Center for Astrophysics, 60 Garden St., Cambridge, MA 02138, USA \\
$^{11}$ Department of Physics, Carnegie Mellon University, 5000 Forbes Ave., Pittsburgh, PA 15213, USA\\
$^{12}$ Department of Physics and Center for Cosmology and Astro-Particle Physics, Ohio State University, USA\\
$^{13}$ University College London, Gower Street, London WC1E 6BT, UK\\
$^{14}$ Institute of Cosmology \& Gravitation, University of Portsmouth, Dennis Sciama Building, Portsmouth PO1 3FX, UK\\
$^{15}$ National Abastumani Astrophysical Observatory, Ilia State University, 2A Kazbegi Ave., GE-1060 Tbilisi, Georgia\\
$^{16}$ Department of Astronomy and Astrophysics, The Pennsylvania State University, University Park, PA 16802, USA\\
$^{17}$ Institute for Gravitation and the Cosmos, The Pennsylvania State University, University Park, PA 16802, USA\\
$^{18}$ Steward Observatory, University of Arizona, 933 N. Cherry Ave., Tucson, AZ 85721, USA\\
$^{19}$ Center for Cosmology and Particle Physics, New York University, New York, NY 10003, USA\\
$^{20}$ Department of Astronomy, University of Wisconsin-Madison, 475 N. Charter Street, Madison, WI 53706, USA\\
$^{21}$ Department of Physical Sciences, The Open University,	Milton Keynes, MK7 6AA, UK\\
$^{22}$ Department of Physics, University of California Berkeley, CA 94720, USA\\
$^{23}$ Department of Astronomy, Case Western Reserve University, Cleveland, OH 44106, USA
}
\date{Submitted to MNRAS}

\maketitle
\clearpage
\begin{abstract}
We explore the cosmological implications of the angle-averaged correlation function, $\xi(s)$, and the 
clustering wedges, $\xi_{\perp}(s)$ and $\xi_{\parallel}(s)$, of the LOWZ and CMASS galaxy samples from
Data Release 10 and 11 of the SDSS-III Baryon Oscillation Spectroscopic Survey. 
Our results show no significant evidence for a deviation from the standard $\Lambda$CDM model.
The combination of the information from our clustering measurements with recent data from the cosmic
microwave background is sufficient to constrain the curvature of the Universe to $\Omega_k =0.0010\pm0.0029$,
the total neutrino mass to $\sum m_\nu< 0.23\,{\rm eV} $ (95\% confidence level), the effective number of relativistic
species to $N_{\rm eff}=3.31\pm0.27$, and the dark energy equation of state to $w_{\rm DE}=-1.051\pm0.076$.
These limits are further improved by adding information from type Ia supernovae and baryon acoustic
oscillations from other samples. In particular, this data set combination is completely consistent with a
time-independent dark energy equation of state, in which case we find $w_{\rm DE}=-1.024\pm0.052$.
We explore the constraints on the growth-rate of cosmic structures assuming $f(z)=\Omega_{\rm m}(z)^\gamma$
and obtain $\gamma=0.69\pm0.15$, in agreement with the predictions from general relativity of $\gamma=0.55$. 
\end{abstract}
\begin{keywords}
cosmological parameters, large scale structure of the universe
\end{keywords}

\pagebreak
\section{Introduction}
\label{sec:intro}

The large-scale distribution of galaxies contains the signature of
acoustic waves that propagated through the Universe prior to the epoch of recombination. 
This signal, referred to as baryon acoustic oscillations (BAO), appears as a modulation in the 
amplitude of the galaxy power spectrum, $P(k)$, and a broad peak in
the large-scale two-point correlation function, $\xi(s)$ \citep{Eisenstein1998,Meiksin1999,Matsubara2004}.
The wavelength of the oscillations in $P(k)$ and the location of the peak in $\xi(s)$  
can be associated with the maximum distance that these acoustic waves can travel before 
the decoupling of matter and radiation, that is, the sound horizon at the drag redshift, $r_{\rm d}$.
As this scale can be constrained with high accuracy from observations of the cosmic
microwave background (CMB), the acoustic scale inferred from the clustering 
of galaxy samples at different redshifts can be used as a standard ruler to measure 
the distance-redshift relation, providing a powerful
and robust probe of the expansion history of the Universe \citep{Blake2003,Linder2003}.

The BAO signal was first detected in the clustering of the 
Two-degree Field Galaxy Redshift survey \citep[2dFGRS,][]{Colless2001,Colless2003}
by \citet{Cole2005} and the luminous red galaxy \citep[LRG,][]{Eisenstein2001} sample  
of the Sloan Digital Sky Survey \citep[SDSS,][]{York2000} by \citet{Eisenstein2005}.
Since then, subsequent analyses on various galaxy samples have provided BAO measurements with increasing
precision \citep{Padmanabhan2007,Beutler2011,Blake2011,Xu2012,Seo2012,Anderson2012,Anderson2013}.
Using these results it is now possible to construct a Hubble diagram based entirely on BAO distance
measurements. It has become standard practice to use this information, in combination with additional
data sets, when deriving constraints on cosmological parameters.

Separate measurements of the acoustic scale in the directions parallel and perpendicular to the 
line of sight can be used to obtain constraints on the Hubble parameter, $H(z)$, and the angular
diameter distance, $D_{\rm A}(z)$, through the Alcock--Paczynski test 
\citep{Alcock1979,Hu2003}. However, the BAO signal on angle-averaged clustering
measurements such as $P(k)$ or $\xi(s)$ provide estimates of the average distance 
$D_{\rm V}(z)\propto D_{\rm A}(z)^2/H(z)$.
Although most analyses have focused on angle-averaged quantities, the large volumes
probed by present-day galaxy samples make it possible to extend these analyses to anisotropic
clustering measurements \citep{Cabre2009,Blake2012,Xu2012,Anderson2013,Kazin2013} 
using the full power of the BAO test.
 
The clustering of galaxies encodes additional information beyond that contained in the BAO signal that
can significantly improve the cosmological constraints derived from large-scale structure (LSS) data sets.
This extra information is particularly important for anisotropic clustering measurements, 
where the signature of the so-called redshift-space distortions (RSD) can be used to constrain the
growth rate of cosmic structures \citep{Guzzo2008}. 
In this way, anisotropic clustering measurements can provide information 
of the expansion history of the Universe and the growth of density
fluctuations, which can be used to distinguish between the dark energy and modified gravity
scenarios for the origin of cosmic acceleration. 

The most accurate BAO measurements to date have been obtained from the Baryon Oscillation
Spectroscopic Survey \citep[BOSS,][]{Dawson2013}, which is one of the four component surveys of
SDSS-III \citep{Eisenstein2011}.
After applying a modified version of the reconstruction technique of \citet{Eisenstein2007b}, 
the BAO signal in the galaxy clustering of BOSS SDSS Data Release 9
\citep[DR9,][]{Ahn2012} provided a 1.7 per cent accuracy measurement of the average distance
$D_{\rm V}(z)$ at $z=0.57$ \citep{Anderson2012}, as well as separate constraints on
$D_{\rm A}(z)$ and $H(z)$ at the same redshift with 3 and 8 per cent accuracy,
respectively \citep{Anderson2013}. 
These measurements have been complemented by analyses of the full shape of isotropic and anisotropic clustering
measurements \citep{Reid2012,Sanchez2012,Sanchez2013,Samushia2013,Chuang2013a}.
Besides galaxy clustering analyses, a sample of high-redshift quasars from BOSS has been used to detect
for the first time the signature of the BAO in the fluctuations of the Lyman-$\alpha$ forest at
$z\simeq2.4$ \citep{Busca2013,Slosar2013}.

\begin{figure*}
\centering
\includegraphics[width=0.47\textwidth]{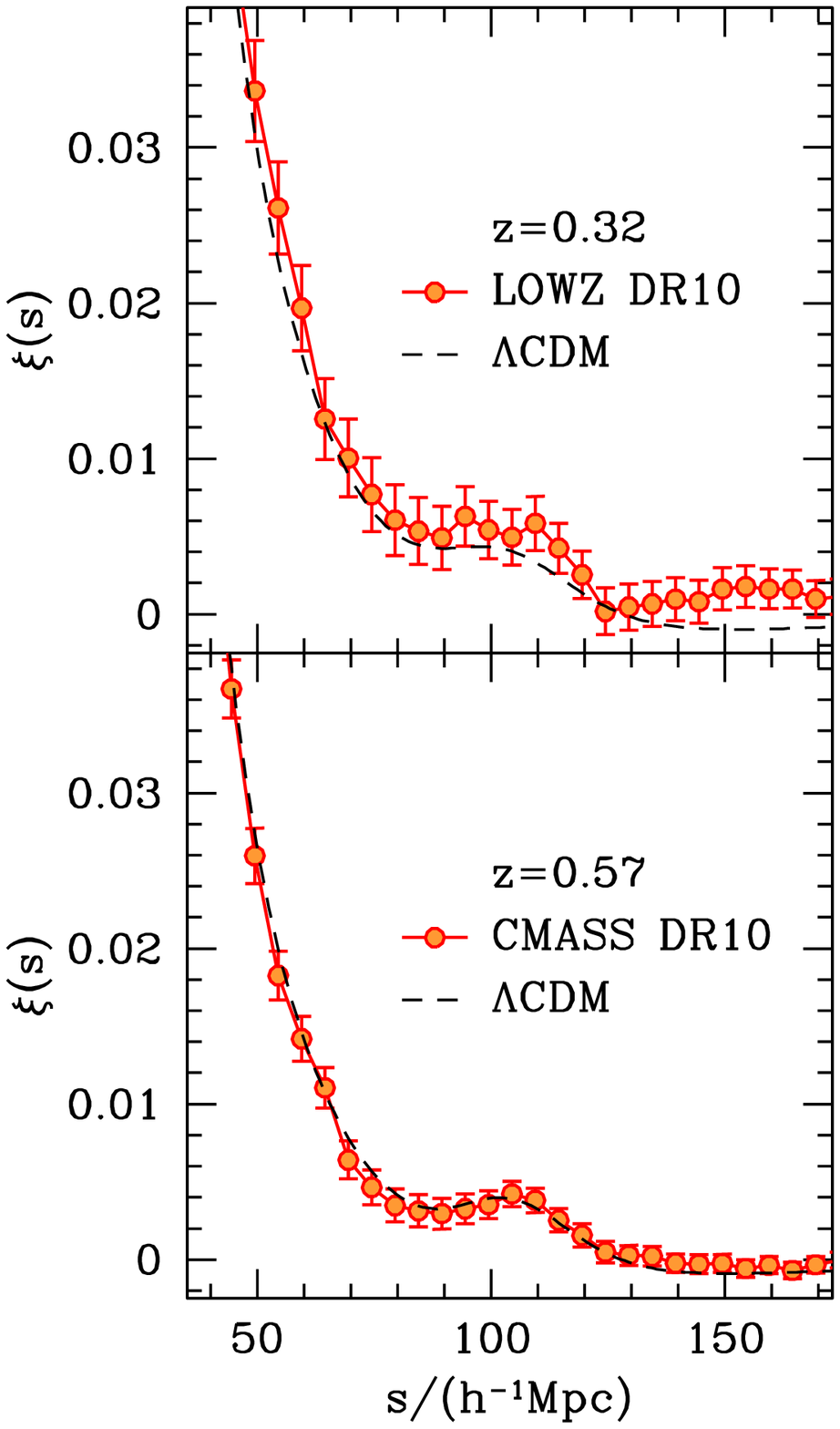}
\includegraphics[width=0.47\textwidth]{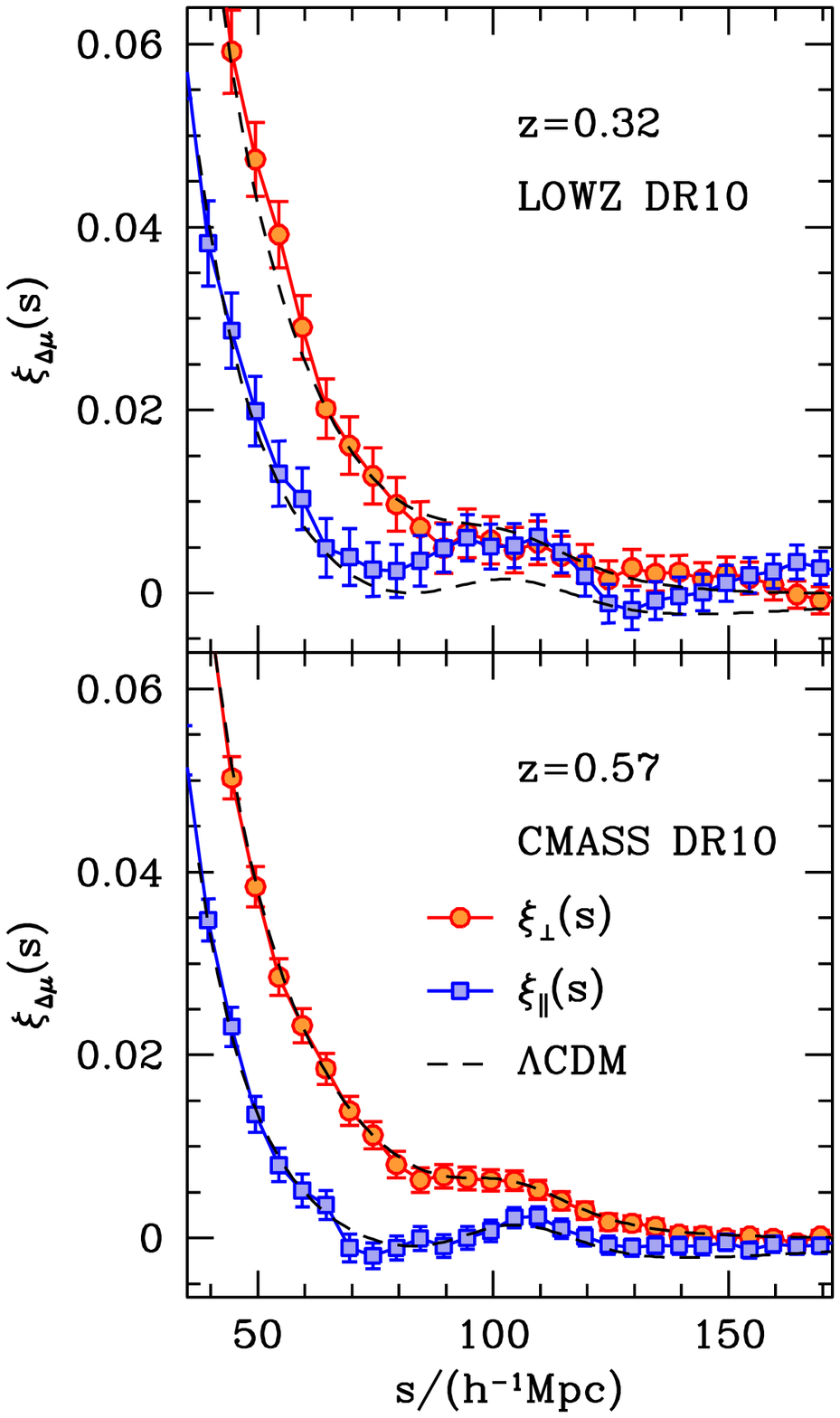}
\caption{
Angle-averaged correlation functions $\xi(s)$, (left panels) and clustering wedges $\xi_{\perp}(s)$ and $\xi_{\parallel}(s)$
(right panels) of the LOWZ and CMASS DR10 galaxy samples. The error bars were derived
from the diagonal entries of the full covariance matrices obtained as described in Sec.~\ref{sec:covariance}. 
The dashed lines correspond to the best-fitting $\Lambda$CDM model obtained
from the combination of information from the full shape of the LOWZ and CMASS DR11 clustering wedges with
the CMB temperature fluctuation measurements from Planck and the nine-year polarization measurements
from WMAP (see Section \ref{sec:lcdm}).
}
\label{fig:measurdr10}
\end{figure*}

In this paper we use information from the full-shape of the two-point correlation function and the
clustering wedges statistic \citep*{Kazin2012} 
measured from BOSS data to derive constraints on cosmological parameters.
We extend the analyses of \citet{Sanchez2012,Sanchez2013} based on a high-redshift galaxy sample
from BOSS DR9 to the data corresponding to DR10 \citep{Ahn2013} and DR11 (internal data-release),
including results from the low-redshift BOSS galaxy sample. 

\begin{figure*}
\centering
\includegraphics[width=0.47\textwidth]{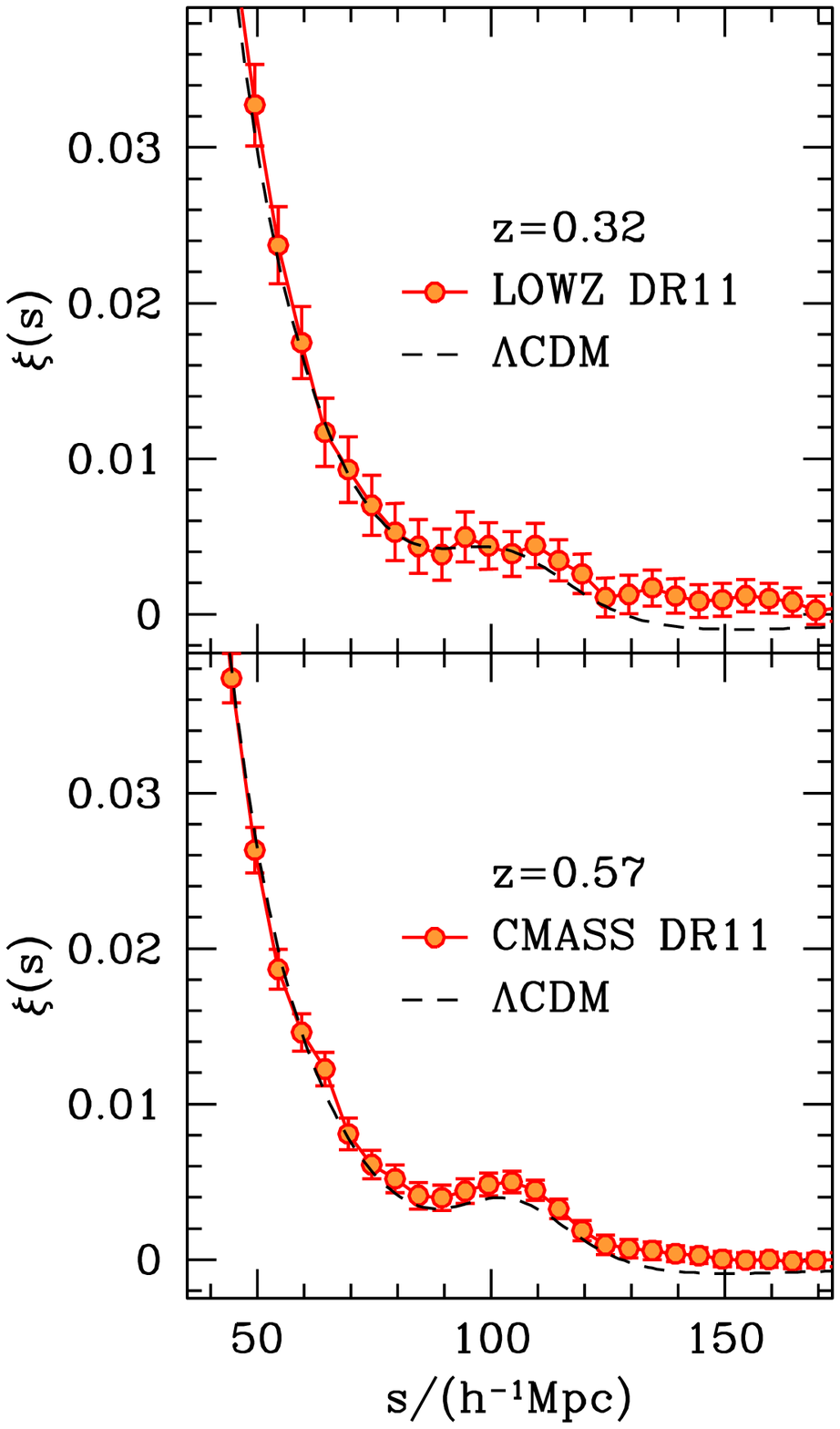}
\includegraphics[width=0.47\textwidth]{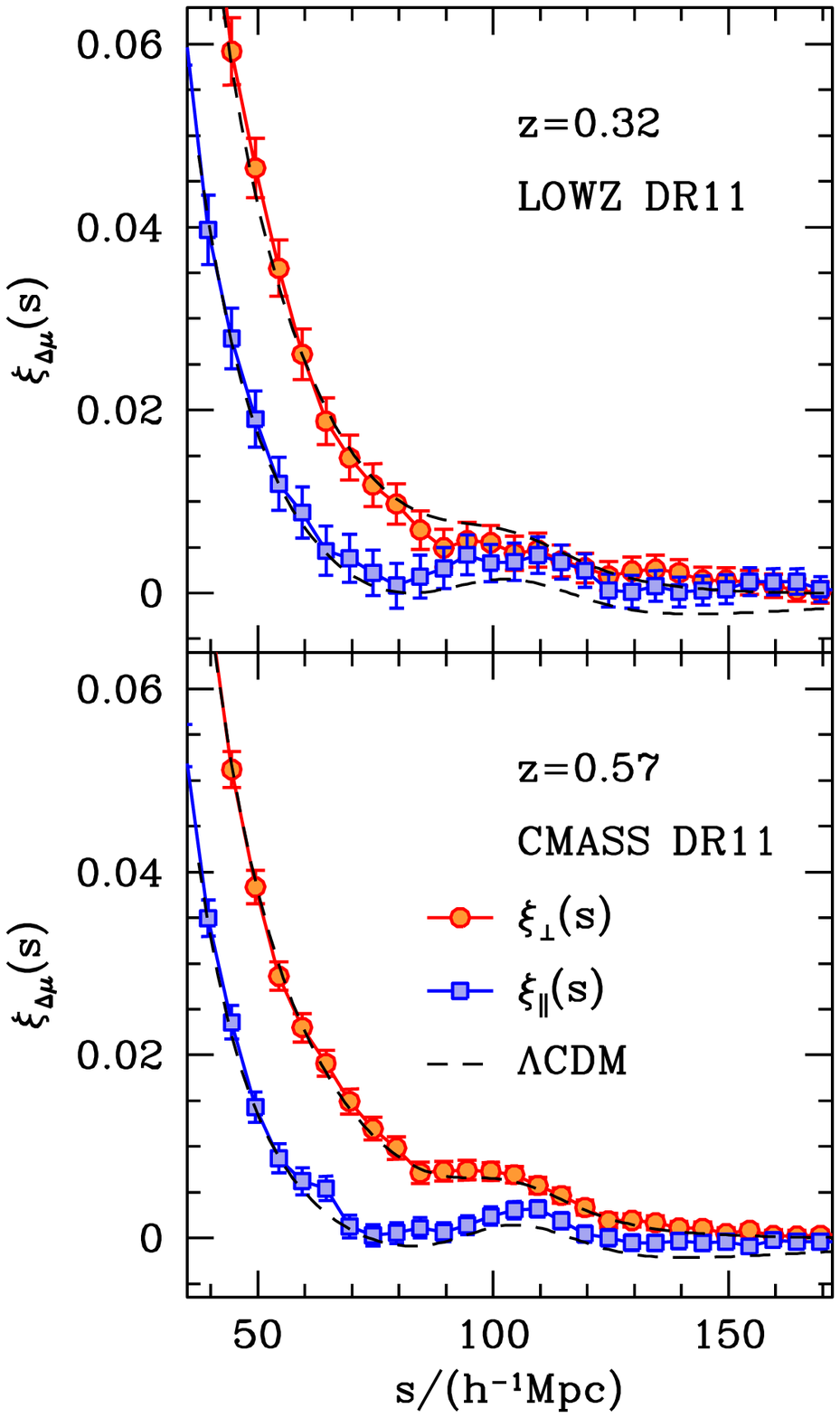}
\caption{
The same as Figure 1, but for the LOWZ and CMASS DR11 galaxy samples.
}
\label{fig:measurdr11}  
\end{figure*}

As the statistical uncertainties characterizing different cosmological observations become smaller, 
it is important to explore potential systematics that can be introduced by the analysis techniques 
and models applied to the data. The comparison of the results obtained by applying multiple methods to the
same data can be used to identify the presence of systematics errors.
Our analysis is part of a series of papers examining the clustering properties of the 
BOSS DR10 and DR11 galaxy samples with different methodologies.
Tojeiro et al. (in preparation) and \citet{Aardwolf2013} analyse the isotropic
and anisotropic BAO signal in these samples and explore their cosmological implications. 
\citet{Ross2013} study the sensitivity of these BAO measurements to the properties of the galaxy
population being analysed. \citet{Vargas2013} investigate the potential systematic errors affecting
anisotropic BAO measurements. \citet{Percival2013} perform a detailed analysis of the effect of
the uncertainties in the covariance matrices determined from mock catalogues on the obtained constraints.
These analyses are complemented by those of \citet{Chuang2013b}, \citet{Samushia2013b} and \citet{Beutler2013},
who analyse the full shape of the monopole-quadrupole pair in configuration and Fourier space.  
These studies attempt to condense the information of the clustering measurements into a few numbers reflecting
the geometric constraints and the measurements of the growth of structures, that are then compared with the 
predictions from different cosmological models. We follow an alternative approach in which we perform the
comparison with cosmological models at the level of the galaxy clustering measurements themselves. 
The consistency of the results presented here and those of our companion papers is a reassuring indication of the 
robustness of our results. 

The outline of this paper is as follows. In Section~\ref{sec:data} we describe our galaxy sample,
the procedure followed to obtain our clustering measurements 
and their respective covariance matrices, as well as the additional data sets included in our analysis.
In Section~\ref{sec:method} we review our model of the full shape of the correlation function and the 
clustering wedges and our methodology to obtain cosmological constraints. In Section~\ref{sec:results} we present
the constraints on cosmological parameters obtained from different combinations of data sets and parameter
spaces. Finally, Section~\ref{sec:conclusions} contains our main conclusions.

\section{The data}
\label{sec:data}

\subsection{The Baryon Oscillation Spectroscopic Survey}
\label{sec:boss}

\subsubsection{Galaxy clustering measurements from BOSS}
\label{sec:clustering}

We use the LOWZ and CMASS samples of BOSS corresponding to SDSS DR10 \citep{Ahn2013} and DR11, which 
will become publicly available with the final data release of the survey.
These galaxy samples were selected on the basis of the SDSS multicolour photometric observations \citep{Gunn1998,Gunn2006}
to cover the redshift range $0.15<z<0.7$ with a roughly uniform comoving number density
$n\simeq3 \times 10^{-4} h^3{\rm Mpc}^{-3}$ \citep[][Padmanabhan et al. in preparation]{Eisenstein2011,Dawson2013}.
Up to $\sim30$ and 2 per cent of LOWZ and CMASS targets, respectively, were observed during the SDSS I/II surveys \citep{York2000}
and thus already have a redshift. The remaining redshifts were measured from the spectra obtained with the
double-armed BOSS spectrographs \citep{Smee2013} by applying the minimum-$\chi^2$ template-fitting procedure
described in \citet{Aihara2011} and \citet{Bolton2012}. 

The LOWZ sample consists primarily of red galaxies that lie in massive haloes, with a satellite fraction of
12 per cent \citep{Parejko2013}. The CMASS sample is approximately complete down to a limiting stellar mass of
$M\simeq10^{11.3}\,{\rm M}_{\odot}$ \citep{Maraston2013},
and has a $\sim$10 per cent satellite fraction \citep{White2011,Nuza2013}.
Although this sample is dominated by early type galaxies, it contains a significant fraction of massive spirals
\citep[$\sim$26 per cent,][]{Masters2011}.
\citet{Aardwolf2013} describes the construction of catalogues for LSS analyses based on these samples.
We use these samples separately, restricting our analysis to the redshift ranges
$0.15 < z < 0.43$ for the LOWZ sample, and $0.43 < z < 0.7$ for the CMASS galaxies. 

We study the clustering properties of these galaxy samples by means of the angle-averaged correlation
function, $\xi(s)$, and the clustering wedges statistic \citep{Kazin2012}, $\xi_{\Delta\mu}(s)$,
which corresponds to the average of the full two-dimensional correlation function $\xi(\mu,s)$ over the
interval $\Delta\mu=\mu_{\rm max}-\mu_{\rm min}$, that is
\begin{equation}
\xi_{\Delta\mu}(s)\equiv \frac{1}{\Delta \mu}\int^{\mu_{\rm max}}_{\mu_{\rm min}}{\xi(\mu,s)}\,{{\rm d}\mu}.
\label{eq:wedges}
\end{equation}
We use two wide clustering wedges, $\xi_{\perp}(s)$ and $\xi_{\parallel}(s)$, defined 
for the intervals $0 \le \mu \le 0.5$ and $0.5 \le \mu \le 1$, respectively. 
The basic procedure implemented to obtain these measurements from the LOWZ and CMASS samples is analogous
to that of \citet{Aardwolf2013} and \citet{Sanchez2013}. Here we summarize the most important
points and refer the reader to these studies for more details.

We convert the observed redshifts into distances assuming a flat $\Lambda$CDM fiducial cosmology
characterized by a matter density parameter of $\Omega_{\rm m}=0.274$.
We use the estimator of \citet{Landy1993} to compute the full correlation function $\xi(\mu,s)$ 
of the LOWZ and CMASS samples, with random samples following the same selection function as the original
catalogues but containing 50 times more objects. The value of $\mu$ of a given pair is defined as
the cosine of the angle between the separation vector, $\mathbf{s}$, and the line-of-sight direction
at the midpoint of $\mathbf{s}$.
We infer the correlation function $\xi(s)$ and the clustering wedges $\xi_{\perp}(s)$ and $\xi_{\parallel}(s)$ 
by averaging the full $\xi(\mu,s)$ over the corresponding $\mu$ intervals.
As discussed in \citet{Kazin2012}, this procedure correctly accounts for the $\mu$ dependence of the random-random 
counts, which is ignored when the estimator of \citet{Landy1993} is applied to the averaged counts directly, leading
to a bias in the recovered clustering measurements.

 When computing the pair counts, we assign a series of weights to each object in our catalogue.
First, we apply a radial weight designed to minimize the variance of our measurements \citep*{Feldman1994} given by 
\begin{equation}
w_{\rm r}=\frac{1}{1+P_{w}\bar{n}(z)},
\label{eq:wradial}
\end{equation}   
where $\bar{n}(z)$ is the expected number density of the catalogue at the given redshift and $P_{w}$ is
a scale-independent parameter, which we set to $P_w=2\times 10^4\,h^{-3}{\rm Mpc}^3$.
We also include angular weights to account for redshift failures and fibre collisions. For the CMASS
sample we apply additional weights to correct for the systematic effect introduced by the local stellar
density and the seeing of the observations, as described in detail in \citet{Aardwolf2013}.

The left panels of Figs.~\ref{fig:measurdr10} and \ref{fig:measurdr11} show the resulting angle-averaged correlation function $\xi(s)$ of the SDSS-DR10 and DR11 LOWZ (upper) and CMASS (bottom) samples, respectively, while the left panels show the 
corresponding clustering wedges $\xi_{\perp}(s)$ (circles) and $\xi_{\parallel}(s)$ (squares). 
The anisotropic clustering pattern generated by redshift-space distortions leads to significant differences
in the amplitude and shape of the two clustering wedges, with $\xi_{\parallel}(s)$ showing a lower amplitude
and a stronger damping of the BAO peak than $\xi_{\perp}(s)$.
The dashed lines correspond to the best-fitting $\Lambda$CDM model obtained from the combination of 
the LOWZ and CMASS DR11 clustering wedges with CMB observations from the Planck satellite 
\citep{Planck2013a} and the CMB polarization measurements from WMAP \citep{Bennett2013}
as described in Section~\ref{sec:lcdm}, which provide an excellent description
of all our measurements.  

\subsubsection{Covariance matrix estimation}
\label{sec:covariance}

When comparing our BOSS clustering measurements with theoretical predictions
we assume a Gaussian likelihood function of the form ${\cal L}\propto\exp(-\chi^2/2)$.
The calculation of the $\chi^2$ value of a given model requires the knowledge of the inverse
covariance matrix of our measurements, which we estimate using  
mock catalogues matching the selection functions of the LOWZ and CMASS samples.
These mocks were constructed from two sets of {\sc PTHalos} realizations \citep{Scoccimarro2002},
corresponding to our fiducial cosmology, as described in
\citet{Manera2013} and Manera et al. (in preparation)\footnote{http://www.marcmanera.net/mocks/}.
Our CMASS mocks are based on 600 independent simulations with a box size of $L_{\rm box}=2.4\,h^{-1}{\rm Gpc}$, 
while those of the LOWZ sample were constructed from a separate set of 500 boxes with the same volume. 
In the construction of these mocks, the Northern Galactic Cap (NGC) and Southern Galactic Cap (SGC)
components of the survey were considered as being independent, and sampled separately from the same
PTHalos realizations.
The volume of the LOWZ sample allowed us to obtain two separate NGC and SGC mocks per
{\sc PTHalos} realization, leading to 1000 independent combined NGC+SGC LOWZ mock catalogues.
The larger volume of the CMASS sample makes it more difficult to construct mocks of the NGC and SGC
components from the boxes without overlap. This means that the NGC and SGC
CMASS mocks drawn from the same box are not independent. For DR10 the overlap 
between the NGC and SGC mocks is approximately 75 per cent
of the area covered by the SGC, while for DR11 the whole of the
Southern component is also covered by the NGC.
To account for this overlap in our covariance matrix estimations 
we construct two sets of 300 independent NGC+SGC CMASS mocks, drawing the
matched components from different boxes.

\begin{table}
\centering
  \caption{
    Correction factors of equation~(\ref{eq:d_invcov}) to account for the bias in the estimation of the
inverse covariance matrix.}
\begin{tabular}{@{}lc@{}}
\hline
Measurement                                                 & $(1-D)$ \\
\hline
DR10 \& DR11 LOWZ $\xi(s)$                  & 0.978   \\[0.5mm]
DR10 \& DR11 LOWZ $\xi_{\Delta \mu}(s)$     & 0.953   \\[0.5mm] 
DR10 CMASS $\xi(s)$                         & 0.955   \\[0.5mm]
DR10 CMASS $\xi_{\Delta \mu}(s)$            & 0.913   \\[0.5mm]
DR11 CMASS $\xi(s)$                         & 0.950   \\[0.5mm]
DR11 CMASS $\xi_{\Delta \mu}(s)$            & 0.902   \\[0.5mm]
\hline
\end{tabular}
\label{tab:dvalues}
\end{table}

We measured the angle-averaged correlation function and the clustering wedges of each LOWZ and
CMASS mock catalogue using the same binning and weighting schemes as for the real data. 
These measurements were used to obtain an estimate of the full covariance matrix $\mathbfss{C}$ 
of our clustering measurements.
For the CMASS sample we define our covariance matrix as the average of the results 
obtained in the two sets of independent mocks. 
The error bars in Figs.~\ref{fig:measurdr10} and \ref{fig:measurdr11} correspond to the square root
of the diagonal entries in $\mathbfss{C}$.

 Our estimations of $\mathbfss{C}$ are affected by noise, as they are inferred from a finite number of
mock catalogues. This uncertainty has important implications on the derived constraints. 
The distribution of covariance matrices recovered from multiple, independent sets of simulations
follows a Wishart distribution, and its inverse, $\mathbfss{C}^{-1}$, an inverse-Wishart distribution
\citep{Wishart1928}.
As the inverse Wishart distribution is asymmetric, $\mathbfss{ C}^{-1}$ provides a biased estimate of the true
inverse covariance matrix \citep[see e.g.][]{Hartlap2007,Taylor2013,Percival2013}.
This bias can be corrected for by rescaling the inverse covariance matrix as
\begin{equation}
\hat{\mathbfss{C}}^{-1}=(1-D)\,\mathbfss{C}^{-1},
\label{eq:cinv}
\end{equation}
with 
\begin{equation}
D=\frac{N_{\rm bins}+1}{N_{\rm mocks}-1},
\label{eq:d_invcov}
\end{equation}
where $N_{\rm mocks}$ is to the total number of mocks used to estimate $\mathbfss{C}$ and $N_{\rm bins}$
corresponds to the total number of bins in our measurements.
We restrict our analysis to $40\,h^{-1}{\rm Mpc}< s < 160 \,h^{-1}{\rm Mpc}$
with a bin-width of $ds=5\,h^{-1}{\rm Mpc}$, leading to $N_{\rm bins}=30$ for the angle-averaged
correlation function and $N_{\rm bins}=60$ for the clustering wedges.
 
Equation~(\ref{eq:d_invcov}) shows that an accurate estimate of the inverse covariance matrix requires
a large number of independent realizations. 
While for the LOWZ sample our covariance matrix estimates are based on $N_{\rm mocks}=1000$, 
for the CMASS sample we use two sets of 300 independent mocks. As these two sets are correlated, 
their combination does not lead to the same noise that would correspond to using 600
independent estimates. To account for this fact we follow \citet{Percival2013} 
and compute the correction term $D$ in equation~(\ref{eq:d_invcov}) using $N_{\rm mocks}=300$, multiplied by
$(1 + r^2)/2$, where $r$ corresponds to the correlation coefficient between the mock clustering measurements.
The volume overlap between our mock catalogues implies 
 $r = 0.33$ for DR10 and $r = 0.49$ for DR11 \citep[for more details see][]{Percival2013}.
The resulting correction factors for the clustering measurements used in our analysis
are listed in Table~\ref{tab:dvalues}.

Although the correction factor of equation~(\ref{eq:d_invcov}) leads to an unbiased estimation of the 
inverse covariance matrix, it does not correct for the effect of the uncertainties in this estimate,
which should be propagated into the obtained cosmological constraints.
\citet{Percival2013} present a detailed description of the effect of the noise in the covariance
matrix estimated from a set of mock realizations and derive formulae for their impact
on the errors of the cosmological constraints measured by integrating over the likelihood function. 
They demonstrated that, to account for this extra uncertainty, the recovered parameter covariance 
constraints must be rescaled by a factor that depends on $N_{\rm bins}$, $N_{\rm mocks}$ 
and the number of parameters included in the analysis, $N_{\rm par}$ \citep[see equation~18 in][]{Percival2013}. 
Depending on the parameter space, our choice of range of scales and binning leads to a modest correction factor
of at most 2.4 per cent for the results inferred from the clustering wedges, which we include in our constraints,
and a negligible correction (less than 0.3 per cent) for the results obtained from the angle-averaged correlation
function.

\subsection{Additional data-sets}
\label{sec:moredata}

We combine the information encoded in the full shape of our clustering measurements with additional observations
 in order to improve the obtained cosmological constraints. Here we give a brief description of
each additional data set. 

We use the low-$\ell$ and high-$\ell$ CMB temperature power spectrum from the one-year data
release of the Planck satellite \citep{Planck2013a} with the low-$\ell$ polarization measurements 
from the nine-year of observations of the WMAP satellite \citep{Bennett2013,Hinshaw2013}.
This data set corresponds to the `Planck+WP' case considered in \citet{PlanckXVI2013}.
For simplicity we refer to this combination simply as `Planck'.
We extend this data set using the high-$\ell$ CMB measurements from the Atacama Cosmology Telescope
\citep[ACT,][]{Das2013} and the South Pole Telescope \citep[SPT,][]{Keisler2011,Story2013,Reichardt2012}.
We refer to this combination as `ePlanck'.
We also explore the constraints obtained by replacing the Planck CMB data by the final nine-year results from
the WMAP satellite \citep{Bennett2013, Hinshaw2013} to test the consistency between these data sets. 
We refer to these measurements as `WMAP9'.

We also include information from distance measurements inferred from the angle-averaged BAO
signal from independent samples.
We use the results of \citet{Beutler2011} from the large-scale correlation function of the
6dF Galaxy Survey \citep[6dFGS,][]{Jones2009} corresponding to $z=0.106$,
and the distance measurements inferred from the Lyman-$\alpha$ forest in BOSS
\citep{Busca2013,Slosar2013,Kirkby2013}, corresponding to $z=2.4$.
These data sets constrain the parameter combination
$D_{\rm V}(z)/r_{\rm d}$, where 
\begin{equation}
D_{\rm V}(z)=\left((1+z)^2D_{\rm A}(z)^2\frac{cz}{H(z)}\right)^{1/3}. 
\label{eq:dv}
\end{equation}
We do not include the results of \citet{Xu2012} based on the final SDSS-II LRG sample as this catalogue is
partially contained in the LOWZ sample used here, or the results from \citet{Blake2011} from the
final WiggleZ Dark Energy Survey \citep{Drinkwater2010} at $z=0.44$, 0.6 and 0.73 due to the overlap of
these data with the CMASS sample. 

Finally, we also use the information from the Union2.1 type Ia supernovae (SN) compilation \citep{Suzuki2012}.
This sample combines 833 SN drawn from 19 different data sets using the scheme of the original Union sample of \citet{Kowalski2008}.
For comparison, in some cases we also present results obtained using the SN compilation of \citet{Conley2011}, which includes the
high-redshift SN from the first three years of the Supernova Legacy Survey (SNLS). When using these data we follow the
recipe of \citet{Conley2011} to take into account systematic errors in our cosmological constraints, which requires the
introduction of two additional nuisance parameters, $\alpha$ and $\beta$, related to the stretch-luminosity and colour-luminosity
relationships. When quoting cosmological constraints based on this sample, the values of these parameters are marginalized over.

With the exception of Section~\ref{sec:distance}, we use our BOSS clustering measurements in combination
with CMB data. Unless stated otherwise, we use the information from our clustering measurements in 
the LOWZ and CMASS samples in combination, and refer to them as `BOSS $\xi(s)$' for the angle-averaged
correlation functions and `BOSS $\xi_{\Delta \mu}(s)$' for the clustering wedges. 
Although the bulk of our analysis is based on our DR11 BOSS clustering measurements as they posses smaller
statistical uncertainties, we test the consistency of these results with the constraints 
inferred using their DR10 counterparts. 
Our tightest constraints are obtained when the additional BAO and Union2.1 SN data are also included in
our analysis. We refer to this case as our `Full' combination. 

\section{Methodology}
\label{sec:method}

\subsection{Modelling of our clustering measurements}
\label{sec:model}

We follow the recipe of \citet{Sanchez2013} to model the full shape of the angle-averaged correlation
function and the clustering wedges. This description takes into account the effects of non-linear evolution,
redshift-space distortions and bias which, if unaccounted for, could introduce systematic errors
in the derived cosmological constraints \citep{Smith2008,Crocce2008,Angulo2008,Sanchez2008}.
Here we summarize the main details of our modelling and refer the reader to section 3 of
\citet{Sanchez2013} for more details.

Both the angle-averaged correlation function and the clustering wedges can be obtained by integrating 
$\xi(\mu,s)$ over different $\mu$-intervals.
This means that a theoretical description of these measurements requires a model
of the anisotropic correlation function.
To obtain this model, it is convenient to decompose $\xi(\mu,s)$ in terms of Legendre polynomials as 
\begin{equation}
 \xi(\mu,s)=\sum_{{\rm even}\ \ell} L_{\ell}(\mu)\xi_{\ell}(s),
\label{eq:expansionxi}
\end{equation}
where the multipoles $\xi_{\ell}(s)$ are given by
\begin{equation}
\xi_\ell(s)\equiv \frac{2\ell+1}{2}\int^1_{-1} L_\ell(\mu)\xi(\mu,s)\,{\rm d}\mu.
\label{eq:multipoles_xi}
\end{equation}
These multipoles are related to those of the two-dimensional power spectrum, $P(\mu,k)$, by
\begin{equation}
\xi_{\ell}(s)\equiv \frac{i^{\ell}}{2\pi^2}\int^{\infty}_{0} P_{\ell}(k) j_{\ell}(ks)\,k^2{\rm d}k,
\label{eq:pl2xil}
\end{equation}
where $j_{\ell}(x)$ is the spherical Bessel function of order $\ell$ \citep{Hamilton1997}. 
We describe $P(\mu,k)$ with a simple parametrization as 	
\begin{equation}
P(\mu,k)=\left(\frac{1}{1+(kf\sigma_{\rm v}\mu)^2}\right)^2(1+\beta\mu^2)^2P_{\rm NL}(k),
\label{eq:p2d}
\end{equation}
where $f\equiv\frac{{\rm d}\ln D }{{\rm d}\ln a }$ is the logarithmic structure growth-rate parameter, 
$\beta=f/b$, and $P_{\rm NL}(k)$ represents the non-linear real-space power spectrum, given by
\begin{equation}
P_{\rm NL}(k) = b^2\left[P_{\rm L}(k)\,{\rm e}^{-(k\sigma_{\rm v})^2} + A_{\rm MC} \,P_{\rm MC}(k)\right]
\label{eq:pnl}
\end{equation}
with $b$, $\sigma_{\rm v}$, and $A_{\rm MC}$ treated as free parameters.
Here $P_{\rm MC}(k)$ is given by 
\begin{equation}
 P_{\rm MC}(k) = \frac{1}{4\pi^{3}} \int \rm d^{3}q \,|F_{\rm2}(\mathbf k-\mathbf q,\mathbf q)|^{2} P(|\mathbf k-\mathbf q|)P(q),
\label{eq:pmc}
\end{equation}
where $F_{2}(\mathbf k,\mathbf q)$ is the standard second order kernel of perturbation theory. 
The parametrization of equation~(\ref{eq:pnl}) is motivated by renormalized perturbation theory \citep[RPT,][]{Crocce2006}
and is the basis of the parametrization of the non-linear correlation function proposed by \citet{Crocce2008}.
This simple recipe provides an accurate description of 
the power spectra and correlation functions measured from N-body simulations
\citep[e.g.][]{Sanchez2008,Montesano2010} and has been applied to the analysis of 
numerous galaxy samples \citep{Sanchez2009,Sanchez2012,Sanchez2013,Montesano2012,Beutler2011,Blake2011}.
The Lorentzian pre-factor in equation~(\ref{eq:p2d}) accounts for the
Finger-of-God effect \citep{Jackson1972} under the assumption of an exponential galaxy
velocity distribution function \citep{Park1994,Cole1995}.

Only a small number of multipoles of $\xi(\mu,s)$ have non-negligible values on large scales. 
We base our description of the full $\xi(\mu,s)$ on the multipoles $\xi_{\ell}(s)$ with $\ell < 4$
of the parametrization of equation~(\ref{eq:p2d}).
\citet{Sanchez2013} showed that discarding contributions from multipoles with $\ell\ge4$, 
this simple recipe provides an accurate description of the full shape of the
angle-averaged correlation function and the clustering wedges $\xi_{\perp}(s)$ and $\xi_{\parallel}(s)$, 
leading to unbiased cosmological constraints.

One additional ingredient must be added to our model before it can be compared with real clustering
measurements. As described in Section~\ref{sec:clustering}, these measurements require the
assumption of a fiducial cosmology to convert the observed redshifts into distances. This choice
must be taken into account in our models.
The relation between the true values of $s$ and $\mu$ characterizing a given
galaxy pair and those measured in the fiducial cosmology can be written as
\citep{Ballinger1996}
\begin{eqnarray}
 s &=& s'\sqrt{\alpha_{\parallel}^2(\mu')^2+\alpha_{\perp}^2(1-(\mu')^2)},\label{eq:sfid}\\
\mu &=& \frac{\alpha_{\parallel}\mu'}{\sqrt{\alpha_{\parallel}^2(\mu')^2+\alpha_{\perp}^2(1-(\mu')^2)}},
\label{eq:mufid}
\end{eqnarray}
where the primes denote the quantities in the fiducial cosmology and the scaling factors are given by
\begin{eqnarray}
 \alpha_{\perp} &=& \frac{D_{\rm A}(z_{\rm m})}{D'_{\rm A}(z_{\rm m})},\\
 \alpha_{\parallel} &=& \frac{H'(z_{\rm m})}{H(z_{\rm m})},
\end{eqnarray}
that is, the ratios of the angular diameter distance and the Hubble parameter evaluated at the mean redshift
of the sample being considered, $z_{\rm m}$.
These relations encode the effect of the fiducial cosmology on our clustering measurements,
as they can be used to transform the integral in equation~(\ref{eq:wedges}) from the
fiducial cosmology space to the true cosmology as
\begin{equation}
\xi'_{\Delta \mu}(s')\equiv \frac{1}{\Delta \mu'}\int^{\mu'_{\rm max}}_{\mu'_{\rm min}}\xi(\mu(\mu',s'),s(\mu',s'))\,{{\rm d}\mu'}.
\label{eq:wedges_fid}
\end{equation}
We use this relation to transform our theoretical predictions of $\xi(s)$ and $\xi_{\Delta \mu}(s)$
to the fiducial cosmology assumed in our BOSS clustering measurements.

\begin{table} 
  \caption{
    Cosmological parameters considered in our analysis.
The upper part lists the parameters of the standard $\Lambda$CDM model while the middle section 
lists a number of possible extensions of this parameter space.
The lower part lists a number of important quantities whose values
can be derived from the first two sets.
}
    \begin{tabular}{ll}
    \hline
 \multicolumn{1}{c}{Parameter} & \multicolumn{1}{c}{Description} \\
\hline
\multicolumn{2}{l}{Basic $\Lambda$CDM parameters}   \\[0.4mm]
 \multirow{2}{*}{$\Theta$} & Angular size of the sound horizon at \\
                        &   recombination \\[0.4mm] 
$\omega_{\rm b}$        &  Physical baryon density \\[0.4mm] 
$\omega_{\rm dm}$       &  Physical dark matter density \\[0.4mm] 
$\tau$                  &  Optical depth to reionization \\[0.4mm] 
$n_{\rm s}$             &  Scalar spectral index$^{a}$ \\[0.4mm]
$A_{\rm s}$             &  Amplitude of the scalar perturbations$^{a}$\\[1.6mm]
\multicolumn{2}{l}{ Additional parameters}    \\[0.4mm]
$\Omega_k$              &  Curvature contribution to energy density \\[0.4mm]
$w_0$                   &  Present-day dark energy equation of state, $w_{\rm DE}$ \\[0.4mm]
 $w_a$                  &  Time-dependence of $w_{\rm DE}$ (assuming \\
                        &  $w_{\rm DE}(a)=w_0+w_a(1-a)$)\\[0.4mm]
$\sum m_\nu$            &  Total sum of the neutrino masses \\[0.4mm]
$N_{\rm eff}$           &  Effective number of relativistic species \\[0.4mm]
$\gamma$                & Power-law index of the structure growth-rate\\
                        & parameter, assuming $f(z)=\Omega_{\rm m}^{\gamma}$  \\[1.6mm]
\multicolumn{2}{l}{ Derived parameters} \\[0.4mm]
$\Omega_{\rm m}$        &  Total matter density \\[0.4mm]
$\Omega_{\rm DE}$       &  Dark energy density \\[0.4mm]
$h$                     &  Dimensionless Hubble parameter \\[0.4mm]
$t_{0}/{\rm Gyr}$       &  Age of the Universe \\[0.4mm] 
 $\sigma_{8}$           &  Linear-theory rms mass fluctuations in spheres\\
                        & of radius $8\,h^{-1}{\rm Mpc}$ \\[0.4mm]
$f_\nu$                 & Dark matter fraction in massive neutrinos \\[0.4mm]
$f(z_{\rm m})$          & Structure growth-rate parameter, $f(z)=\frac{d \ln D}{ d\ln a}$ \\
\hline
\end{tabular}
$^{a}$Quoted at the pivot wavenumber of $k_0= 0.05\,{\rm Mpc}^{-1}$.
\label{tab:parameters}
\end{table}

\subsection{Cosmological parameter spaces}
\label{sec:cospar}

\begin{figure*}
\includegraphics[width=0.45\textwidth]{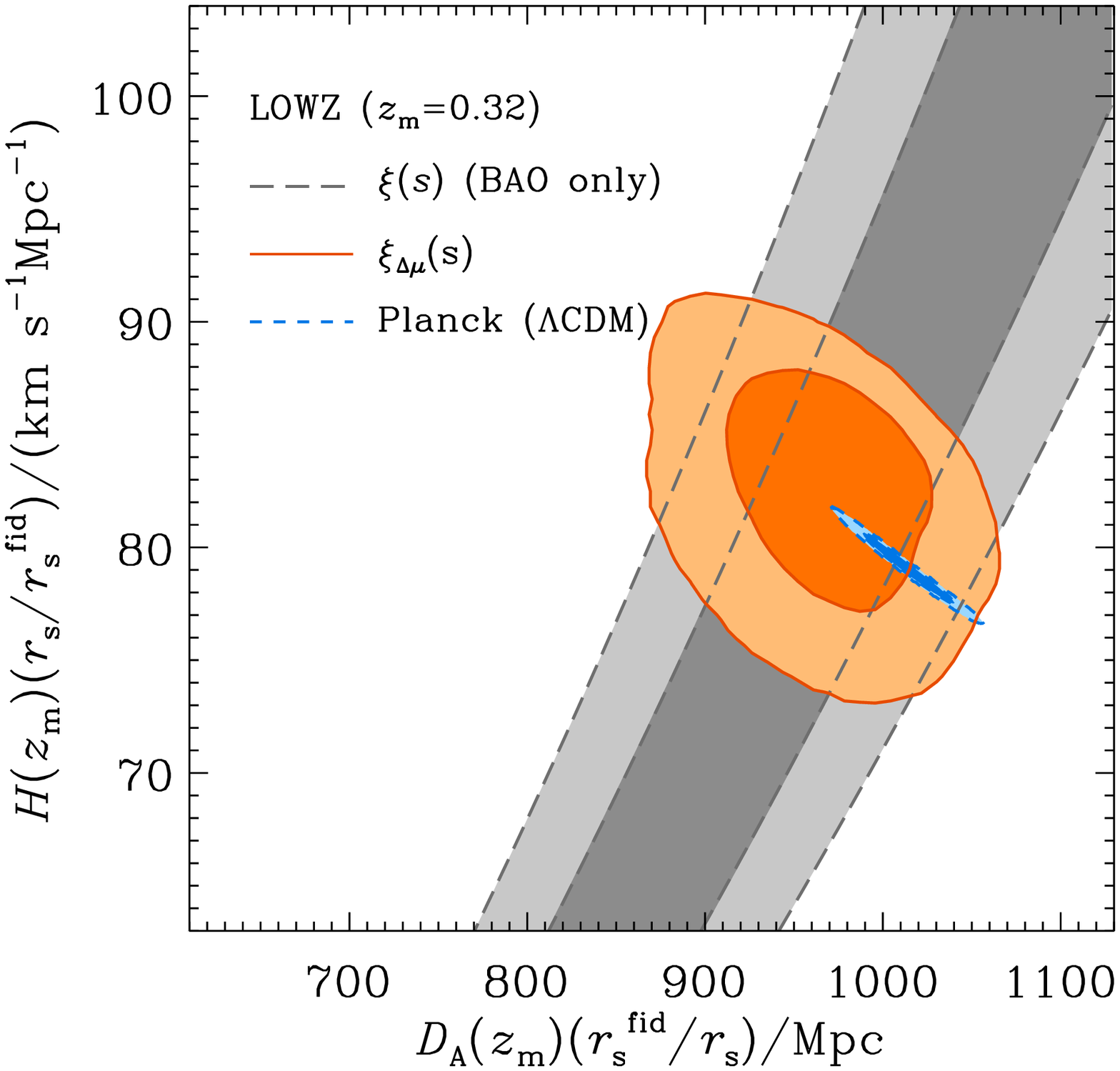}
\includegraphics[width=0.45\textwidth]{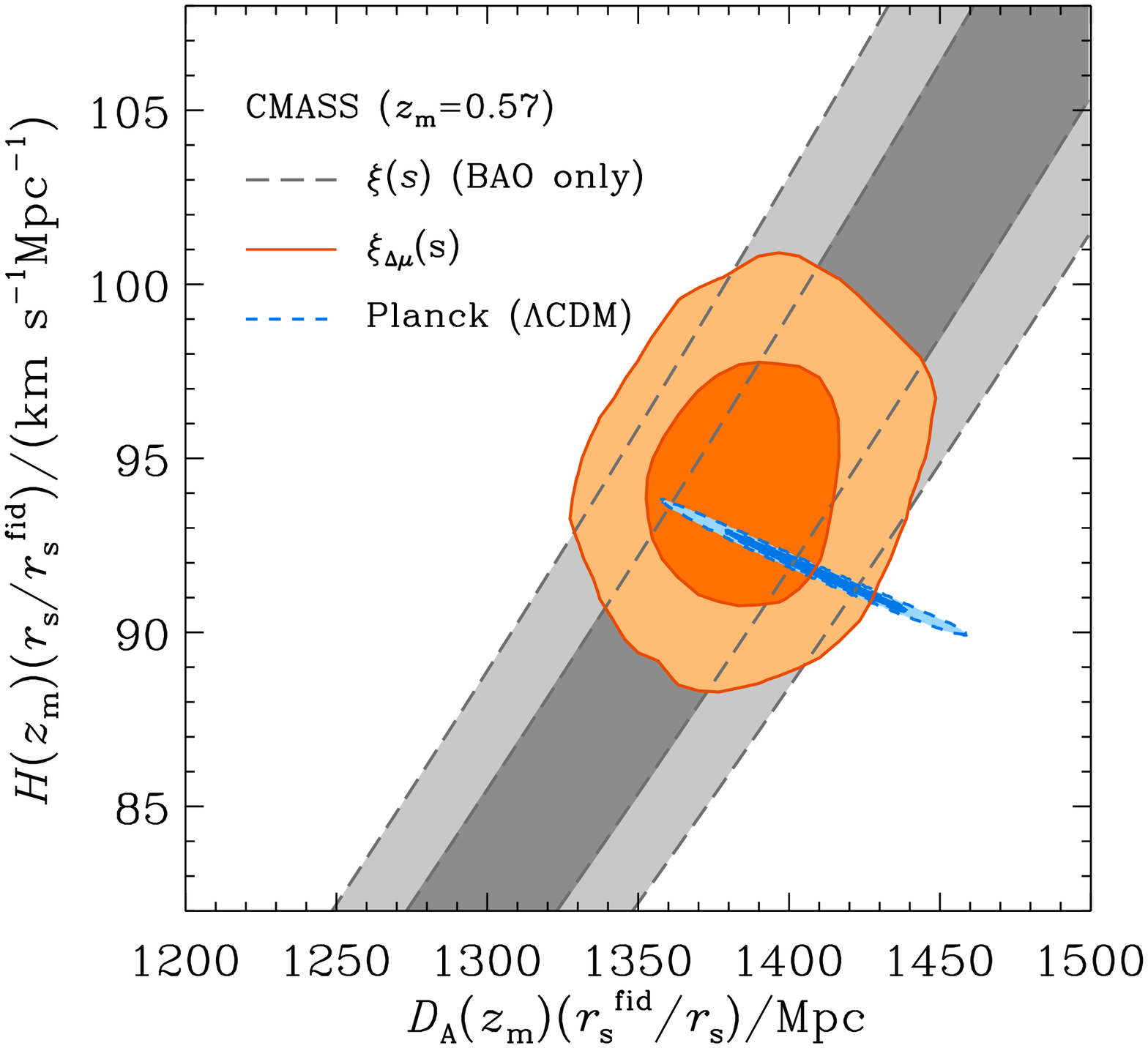}
\caption{
Two-dimensional marginalized constraints in the
$D_{\rm A}(z_{\rm m})\left(r_{\rm d}^{\rm fid}/r_{\rm d}\right)$--$H(z_{\rm m})\left(r_{\rm d}/r_{\rm d}^{\rm fid}\right)$
plane at $z_{\rm m}=0.32$ (left panel) and $z_{\rm m}=0.57$ (right panel) derived from the LOWZ and CMASS DR11 samples, respectively.
The grey long-dashed contours show the results obtained using information from the angle-averaged
correlation function while the red solid lines correspond to those inferred from the clustering wedges $\xi_{\perp}(s)$ and $\xi_{\parallel}(s)$
assuming that $f(z)$ follows the predictions of GR.
The short-dashed contours correspond to the prediction for these parameters derived from the ePlanck data
set (see Section~\ref{sec:moredata}) under the assumption of a $\Lambda$CDM model.
}
\label{fig:dist} 
\end{figure*}

We assume that primordial fluctuations are adiabatic, Gaussian, and have
a power-law spectra of Fourier amplitudes, with a negligible tensor component. 
Table~\ref{tab:parameters} lists the parameters that specify a given cosmological model under these assumptions.
We use the data sets described in Section~\ref{sec:data} to obtain constraints on these parameters.
We start our analysis with the basic $\Lambda$CDM parameter space, which corresponds to a
flat universe where the energy budget contains contributions from baryons, cold dark matter (CDM) and
dark energy, described by an equation of state $w_{\rm DE} =p_{\rm DE}/\rho_{\rm DE}= -1$.
We follow \citet{PlanckXVI2013}
and assume a non-zero fraction of massive neutrinos with a 
fixed total mass $\sum m_\nu=0.06\,$eV. The free parameters required to characterize this model
are listed in the upper part of Table~\ref{tab:parameters}. 
We also explore a number of possible extensions of this parameter
space by allowing for variations on the additional parameters presented in the middle section of
Table~\ref{tab:parameters}. These extensions include more general dark energy models, non-flat universes,
different contributions from massive neutrinos, additional relativistic species and possible deviations
from the predictions of general relativity (GR). 
The final part of Table~\ref{tab:parameters} lists a number of important quantities whose values
can be derived from the remaining parameters.

When studying the properties of the dark energy component we explore the cases of a time-independent
dark energy equation of state $w_{\rm DE}$, and when this parameter is allowed to vary with time, in 
which case we assume the standard parametrization of \citet{Chevallier2001} and \citet{Linder2003} given by
\begin{equation}
w_{\rm DE}(a) = w_0 + w_{a}(1-a). 
\label{eq:wa}
\end{equation}
When exploring the constraints on other potential extensions of the $\Lambda$CDM model, we investigate the
impact of allowing also for variations on $w_{\rm DE}$.

We explore these parameter spaces by means of the Markov Chain Monte Carlo (MCMC) technique. We use 
the June 2013 version of {\sc CosmoMC} \citep{Lewis2002}, modified to include our BOSS
clustering measurements as additional data sets.
This code uses {\sc camb} to compute power spectra for the CMB and matter fluctuations \citep{Lewis2000}, 
which implements the parametrized post-Friedman framework \citep{Hu2007} to account for 
models with $w_{\rm DE}<-1$ and dynamical dark energy models, as described in \citet{Fang2008}. 
Besides the cosmological parameters described here, the analysis of the CMB data requires 
the inclusion of a number of nuisance parameters that are included in our MCMC and marginalized over,
as described in \citet{PlanckXVI2013}. When including clustering measurements from BOSS in our analysis, 
the parameters $b$, $\sigma_{\rm v}$ and $A_{\rm MC}$ for each data set are included as additional
free parameters in our MCMC and marginalized over.

\subsection{The cosmological information in the correlation function and the clustering wedges}
\label{sec:distance}

In this section we analyse the information on geometrical quantities encoded in our measurements of
$\xi(s)$ and the clustering wedges $\xi_{\perp}(s)$ and $\xi_{\parallel}(s)$.
As discussed in detail by \citet{Kazin2012}, while angle-averaged measurements such as $\xi(s)$ provide
constraints on the ratio $D_{\rm V}(z_{\rm m})/r_{\rm d}$,
anisotropic clustering measurement such as the clustering wedges constrain the combinations
$D_{\rm A}(z_{\rm m})/r_{\rm d}$ and $H(z_{\rm m})r_{\rm d}$. 
In this way, the BAO signal in the clustering wedges can be used to break the degeneracy between 
$D_{\rm A}(z_{\rm m})$ and $H(z_{\rm m})$ obtained from $\xi(s)$.
When the full shape of these measurements is taken into account, the extra information 
on $f(z_{\rm m})$ provided by the amplitude difference between $\xi_{\perp}(s)$ and $\xi_{\parallel}(s)$
improves the constraints from those recovered when only the BAO
signal is considered \citep{Sanchez2013}. 

We now investigate the constraints on these quantities that can be derived from our clustering measurements. 
To do this, we explore our most general parameter space treating these quantities as derived parameters, with
their values computed in the context of the cosmological model being tested.
This approach differs from the one applied in our companion papers \citep[Tojeiro et al. in preparation]{Aardwolf2013,Chuang2013b,Samushia2013b,Beutler2013}
where the values of $D_{\rm A}(z)$ and $H(z)$ are treated as free parameters (i.e. without
adopting a specific relation between their values). While this approach will lead to more general
constraints on these parameters than the ones derived here, our results can be used as an indication
of the information content in our clustering measurements and a consistency test with the results of
these analyses. For this exercise, we apply flat priors on the parameters 
$\Phi=(\omega_{\rm b},\omega_{\rm c}, n_{\rm s})$,
which determine the shape of the linear-theory matter power spectrum, centred on the
values corresponding to the best-fitting $\Lambda$CDM model to the Planck CMB data, with a width equivalent
to six times their 68 per cent confidence levels (CL).

Table~\ref{tab:dist} lists the geometrical constraints obtained from the full shape of our clustering
measurements alone, assuming that $f(z_{\rm m})$ follows the predictions of GR. 
Here we have rescaled our results by the sound horizon at the drag redshift for our fiducial cosmology,
$r_{\rm d}^{\rm fid}=149.31\,{\rm Mpc}$, to express them in units of Mpc and 
${\rm km}\,{\rm s}^{-1}{\rm Mpc}^{-1}$.
In all cases the results recovered from the DR10 and DR11 samples are in good agreement.
For the CMASS sample, the extra volume of DR11 leads to a reduction of approximately 20 per cent
in the allowed regions. For the DR11 LOWZ sample, although the limits on $H(z_{\rm m})$ and
$D_{\rm A}(z_{\rm m})$ inferred from the clustering wedges exhibit an improvement of about 10 per cent
with respect to the corresponding DR10 results, the constraints on $D_{\rm V}(z_{\rm m})$ obtained from
$\xi(s)$ show the opposite behaviour. This is consistent with the results of the BAO-only analysis of
Tojeiro et al. (in preparation) and might be related to the higher amplitude of the
acoustic peak in the DR10 LOWZ $\xi(s)$, which leads to a more accurate determination of its centroid.   
Fig.~\ref{fig:dist} shows the two-dimensional marginalized constraints on
$D_{\rm A}(z_{\rm m})\left(r_{\rm d}^{\rm fid}/r_{\rm d}\right)$
and $H(z_{\rm m})\left(r_{\rm d}/r_{\rm d}^{\rm fid}\right)$ at $z_{\rm m}=0.32$
(left panel) and $z_{\rm m}=0.57$ (right panel), derived from the DR11 LOWZ and CMASS samples, respectively. 
The long-dashed lines correspond to the constraints derived from the
angle-averaged correlation function, which correspond to a degeneracy of constant
$D_{\rm V}(z_{\rm m})/r_{\rm d}$. For each sample, the extra information contained in the clustering wedges
breaks this degeneracy, leading to separate constraints on $D_{\rm A}$ and $H$ shown by the solid lines. 

\begin{table} 
\centering
  \caption{Marginalized 68 per cent geometrical constraints derived from the full shape of our clustering
measurements alone, under the assumption that $f(z_{\rm m})$ follows the predictions of GR. We have rescaled
our results by the sound horizon at the drag redshift for our fiducial cosmology,
$r_{\rm d}^{\rm fid}=149.31\,{\rm Mpc}$, to express them in units of Mpc and ${\rm km}\,{\rm s}^{-1}{\rm Mpc}^{-1}$.
}
    \begin{tabular}{@{}lccc@{}}
    \hline
 Data set       &  $D_{\rm V}(z)\left(\frac{r_{\rm d}^{\rm fid}}{r_{\rm d}}\right)$ & 
                 $D_{\rm A}(z)\left(\frac{r_{\rm d}^{\rm fid}}{r_{\rm d}}\right)$ & 
                 $H(z)\left(\frac{r_{\rm d}}{r_{\rm d}^{\rm fid}}\right)$  \\  

\hline
\multicolumn{3}{l}{CMASS ($z_{\rm m}=0.57$)} &    \\[0.4mm]
DR11 $\xi(s)$               &  $2054\pm25$         &  -                 &  - \\[0.4mm]
DR11 $\xi_{\Delta \mu}(s)$  &  $2048\pm25$         & $1387\pm22$        &  $94.3\pm2.4$ \\[0.4mm]
DR10  $\xi(s)$              &  $2046\pm34$         &  -                 &  -   \\[0.4mm]
DR10  $\xi_{\Delta \mu}(s)$ &  $2034\pm31$         & $1385\pm28$        &  $96.0\pm3.4$ \\[1.6mm]
\multicolumn{3}{l}{LOWZ ($z_{\rm m}=0.32$)} &    \\[0.4mm] 
DR11 $\xi(s)$               &  $1254\pm56$         &  -                  &  -   \\[0.4mm]
DR11 $\xi_{\Delta \mu}(s)$  &  $1237\pm42$         & $965\pm37$          & $82.5\pm3.5$ \\[0.4mm]
DR10  $\xi(s)$              &  $1266\pm48$         &  -                  &  -   \\[0.4mm]
DR10 $\xi_{\Delta \mu}(s)$  &  $1237\pm42$         & $960\pm34$          & $81.6\pm3.9$ \\[0.4mm]
\hline
\end{tabular}
\label{tab:dist}
\end{table}

\begin{table} 
\centering
  \caption{Marginalized 68 per cent constraints on geometrical quantities and the growth of structure
derived from the full shape of our clustering wedges, when the assumption that $f(z_{\rm m})$ follows
the predictions of GR is relaxed. We have rescaled our results by the sound horizon at the drag redshift
for our fiducial cosmology, $r_{\rm d}^{\rm fid}=149.31\,{\rm Mpc}$, to express them in units of Mpc and
${\rm km}\,{\rm s}^{-1}{\rm Mpc}^{-1}$.
}
    \begin{tabular}{@{}lccc@{}}
    \hline
 Data set       & $D_{\rm A}(z)\left(\frac{r_{\rm d}^{\rm fid}}{r_{\rm d}}\right)$ & 
                 $H(z)\left(\frac{r_{\rm d}}{r_{\rm d}^{\rm fid}}\right)$  &
                 $f(z)\sigma_8(z)$\\  
\hline
\multicolumn{3}{l}{CMASS ($z_{\rm m}=0.57$)}  &    \\[0.4mm]
DR11 $\xi_{\Delta \mu}(s)$    & $1382\pm26$   &  $93.5\pm3.0$          & $0.417\pm0.045$  \\[0.4mm]
DR10 $\xi_{\Delta \mu}(s)$    & $1381\pm31$   &  $95.5_{-3.2}^{+3.7}$  & $0.469\pm0.060$  \\[1.6mm]
\multicolumn{3}{l}{LOWZ ($z_{\rm m}=0.32$)}   &    \\[0.4mm] 
DR11 $\xi_{\Delta \mu}(s)$    & $965\pm42$    &  $81.7_{-4.4}^{+4.0}$  & $0.48\pm0.10$  \\[0.4mm]
DR10 $\xi_{\Delta \mu}(s)$    & $951\pm39$    &  $80.4\pm3.2$          & $0.43\pm0.10$  \\[0.4mm]
\hline
\end{tabular}
\label{tab:fsig8}
\end{table}

\begin{figure}
\includegraphics[width=0.47\textwidth]{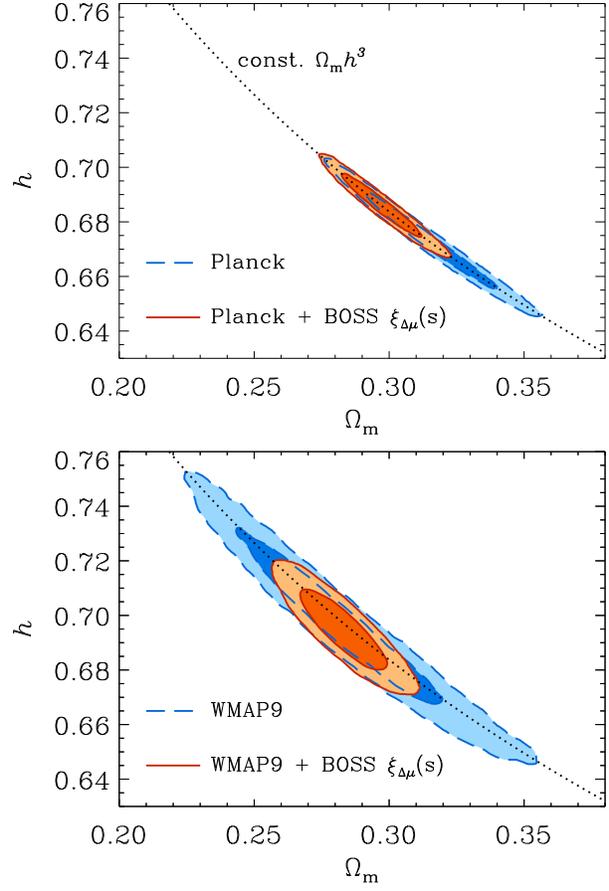}
\caption{
Two-dimensional marginalized constraints in the $\Omega_{\rm m}$--$h$ plane.
The blue dashed lines correspond to the constraints from the Planck (upper panel) and WMAP9
(lower panel) CMB measurements, which follow a degeneracy of constant $\Omega_{\rm m}h^3$, as
indicated by the dotted lines.
The red solid lines show the results obtained when these measurements are combined with the
information from the LOWZ and CMASS DR11 clustering wedges.
}
\label{fig:om_h} 
\end{figure}

We also test the effect of relaxing the assumption of GR to compute $f(z_{\rm m})$ by treating its value as
a free parameter. Table~\ref{tab:fsig8} summarizes the results obtained from the clustering
wedges of our galaxy samples, which are in good agreement with the ones obtained from the BAO-only analysis
of the same galaxy samples in \citet{Aardwolf2013} and Tojeiro et al. (in preparation). 
Exploring this parameter space we can constrain the combination $f\sigma_8(z)$, for which we
obtain $f\sigma_8(0.32)=0.48\pm0.10$ and $f\sigma_8(0.57)=0.417\pm0.045$ using the information of the 
DR11 LOWZ and CMASS galaxy samples, respectively. The constraint derived from the DR11 CMASS sample is in
excellent agreement with the results of \citet{Beutler2013} and \citet{Samushia2013b}, who found 
$f\sigma_8(0.57)=0.419\pm0.042$ and $f\sigma_8(0.57)=0.441\pm0.044$, respectively.
\citet{Chuang2013b} use the information from the multipoles of the LOWZ and CMASS
correlation functions for scales $56 \le s/(h^{-1}{\rm Mpc})\le200$ and find 
$f\sigma_8(0.32)=0.384\pm0.095$ and $f\sigma_8(0.57)=0.354\pm0.059$. Although these values are
lower than the ones reported here, \citet{Chuang2013b} apply a significantly wider prior on the
parameters $\Phi$ and find evidence for an increase in the 
recovered value for this quantity when smaller scales are included in the analysis. 

The dashed lines in Fig.~\ref{fig:dist} correspond to the constraints obtained from 
the ePlanck CMB measurements under the assumption of a $\Lambda$CDM model.
The results obtained from the clustering wedges are in good agreement with the predictions
of the $\Lambda$CDM model that best describes these CMB data, indicating the consistency between
these data sets and their agreement with the $\Lambda$CDM model.
For more general parameter spaces, the region in the 
$D_{\rm A}(z_{\rm m})\left(r_{\rm d}^{\rm fid}/r_{\rm d}\right)$--$H(z_{\rm m})\left(r_{\rm d}/r_{\rm d}^{\rm fid}\right)$ plane 
allowed by the CMB data increases substantially. In these cases the combination of the CMB data
with the information provided by our clustering measurements improves the constraints 
over those recovered from the CMB information alone. 
In Section~\ref{sec:results} we explore the cosmological implications of the information contained in our
clustering measurements.

\begin{table*}  
\centering
  \caption{
    Marginalized 68\% constraints on the most relevant cosmological parameters of the parameter spaces
analysed in Sections~\ref{sec:lcdm} to \ref{sec:fg}, obtained using different combinations of the data sets described
in Section~\ref{sec:data}. A complete list of the constrains obtained in each case can be found in Appendix~\ref{sec:tables}.
}
    \begin{tabular}{@{}lccc@{}}
    \hline
 &\multirow{2}{*}{ePlanck+BOSS $\xi(s)$}  & \multirow{2}{*}{ePlanck+BOSS $\xi_{\Delta \mu}(s)$} & ePlanck + BOSS $\xi_{\Delta \mu}(s)$  \\
                                         &                       &                             &    +BAO+SN         \\  
\hline
\multicolumn{3}{l}{The $\Lambda$CDM model} &  \\[0.6mm]
$h$                   & $0.6824_{-0.0072}^{+0.0072}$ &  $0.6863\pm0.0075$          &  $0.6899\pm0.0070$ \\[0.4mm]
$100\Omega_{\rm m}$   & $30.22_{-0.96}^{+0.94}$      &  $29.71_{-0.96}^{+0.97}$    &  $29.24\pm0.86$   \\[0.4mm]
\hline
\multicolumn{3}{l}{Constant dark energy equation of state} &   \\[0.6mm]
$w_{\rm DE}$            &$-1.31_{-0.16}^{+0.21}$     &  $-1.051\pm0.076$    &  $-1.024\pm0.052$ \\[0.4mm]
$100\Omega_{\rm m}$     & $24.9_{-2.6}^{+3.4}$       &  $28.8\pm1.6$        &  $29.3\pm1.1$   \\[0.4mm]
\hline
\multicolumn{3}{l}{Time-dependent dark energy equation of state} &  \\[0.6mm]
$w_0$                    & $-1.29_{-0.46}^{+0.48}$   & $-0.83_{-0.34}^{+0.38}$     & $-0.95\pm0.14$ \\[0.4mm]
$w_a$                    & $-0.0_{-1.1}^{+1.0}$      & $-0.61_{-0.96}^{+0.89}$        & $-0.29\pm0.47$ \\[0.4mm]
$100\Omega_{\rm m}$      & $25.2_{-6.6}^{+5.7}$      & $30.9_{-3.6}^{+4.1}$        &  $29.5\pm1.3$  \\
\hline                                
\multicolumn{2}{l}{Non-flat models}& &  \\[0.6mm]
$100\Omega_k$             & $0.07\pm0.31$   &  $0.10\pm0.29$             &  $0.15\pm0.29$ \\[0.4mm]
$100\Omega_{\rm m}$       & $30.18\pm0.96$  &  $29.60_{-0.97}^{+0.99}$   &  $29.11\pm0.91$  \\
\hline
\multicolumn{2}{l}{Curvature and dark energy}& &  \\[0.6mm]
$w_{\rm DE}$              & $-1.53_{-0.28}^{+0.24}$  &   $-1.05\pm0.11$           &  $-1.009_{-0.060}^{+0.062}$ \\[0.4mm]
$100\Omega_k$             & $-0.38_{-0.28}^{+0.24}$  &   $0.02\pm0.43$            &  $-0.14\pm0.33$ \\[0.4mm]
$100\Omega_{\rm m}$       & $22.0_{-4.9}^{+3.2}$     &   $28.9\pm2.0$             &  $29.4\pm1.2$   \\
\hline                        
\multicolumn{2}{l}{Massive neutrinos}& &   \\[0.6mm]
$\sum m_\nu$             &  $ < 0.23\,{\rm eV} $ (95\% CL) & $< 0.24\,{\rm eV} $ (95\% CL)   & $ < 0.23\,{\rm eV}$ (95\% CL)\\ 
$f_\nu$                  &  $< 0.017 $ (95\% CL)           & $< 0.019$ (95\% CL)             & $ < 0.017$ (95\% CL) \\[0.4mm]
\hline
\multicolumn{2}{l}{Massive neutrinos and dark energy     }& &  \\[0.6mm]
$\sum m_\nu$            &  $ < 0.49\,{\rm eV} $ (95\% CL) &  $ < 0.47\,{\rm eV} $ (95\% CL)  & $ < 0.33\,{\rm eV}$ (95\% CL)\\ [0.4mm]
$w_{\rm DE}$            &  $-1.49_{-0.30}^{+0.24}$        &  $-1.13\pm0.12$         & $-1.046\pm0.063$ \\
\hline
\multicolumn{2}{l}{Additional relativistic degrees of freedom}& &  \\[0.6mm]
$N_{\rm eff}$            & $3.35\pm0.27$                 &   $3.31\pm0.27$  &   $3.30\pm0.27$ \\ [0.4mm]
$100\Omega_{\rm m}$      & $29.7\pm1.0$                  &   $29.2\pm1.1$   &  $29.1\pm1.0$   \\
\hline
\multicolumn{2}{l}{Deviations from general relativity}& &  \\[0.6mm]
$\gamma$               & -                  &  $0.69\pm0.15$   &   $0.69\pm0.15$ \\ [0.4mm]
$100\Omega_{\rm m}$    & -                  &  $29.76_{-0.90}^{+0.93}$  &  $29.62\pm0.89$  \\
\hline
\multicolumn{2}{l}{Dark energy and modified gravity}& &  \\[0.6mm]
$\gamma$            & -                  &   $0.88\pm0.22$   &   $0.75\pm0.17$ \\[0.4mm]
$w_{\rm DE}$        & -                  &   $-1.15\pm0.11$  &   $-1.055\pm0.057$ \\
\hline
\end{tabular}
\label{tab:extra}
\end{table*}

\section{Cosmological constraints}
\label{sec:results}

Here we describe the cosmological implications of our BOSS clustering measurements.
Section~\ref{sec:lcdm} presents the constraints on the parameters of the standard $\Lambda$CDM model,
while Sections~\ref{sec:wde}-\ref{sec:fg} explore the results obtained in more general parameter
spaces. We pay particular attention to the constraints on the properties of the dark energy component
and study how the limits in other parameters are changed when more general dark energy models are considered.
Appendix~\ref{sec:tables} gives a complete list of the cosmological constraints derived from different
data set combinations, while Table~\ref{tab:extra} summarizes the results on the most important
parameters for the various cases we consider.

\subsection{The $\Lambda$CDM parameter space}
\label{sec:lcdm}

\begin{figure*}
\includegraphics[width=0.45\textwidth]{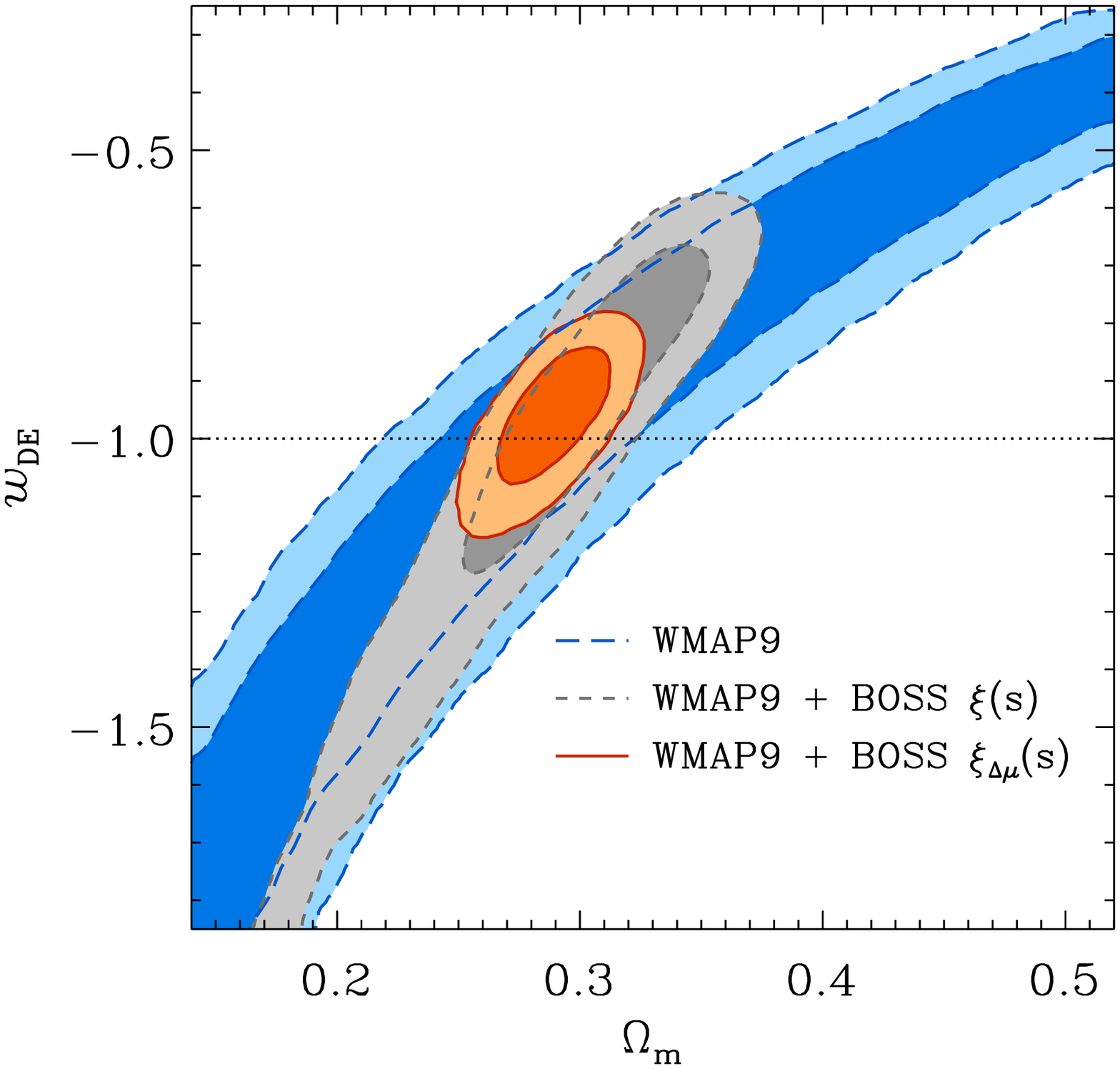}
\includegraphics[width=0.45\textwidth]{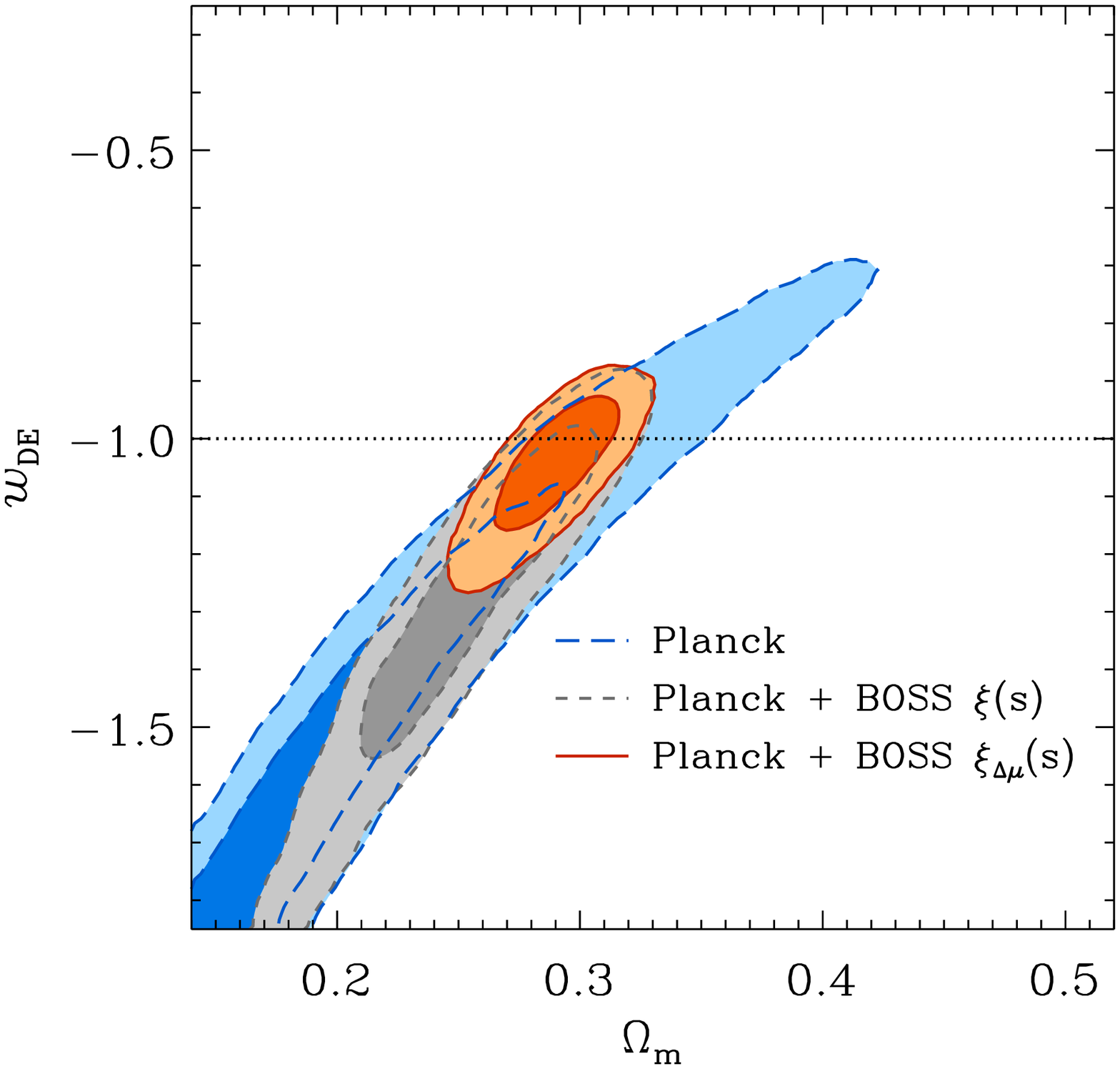}
\caption{
Left panel: marginalized 68 and 95 per cent CL in the $\Omega_{\rm m}$--$w_{\rm DE}$ plane 
for the $\Lambda$CDM parameter set extended by including the redshift-independent value of $w_{\rm DE}$
as an additional parameter. The contours correspond to the results obtained using the
WMAP9-only (blue long-dashed lines), the WMAP9+BOSS $\xi(s)$ combination (grey short-dashed lines) and  
the WMAP9+BOSS $\xi_{\Delta \mu}(s)$ case (red solid lines). The right panel shows the results obtained
when the WMAP9 measurements are replaced by the Planck CMB data set.
The dotted line in both panels corresponds to the $\Lambda$CDM model value of $w_{\rm DE}=-1$.
}
\label{fig:wde} 
\end{figure*}

The simple $\Lambda$CDM model is able to describe an ever increasing amount of precise cosmological
observations, with the CMB measurements from the Planck satellite being perhaps the most striking
example \citep{PlanckXVI2013}. 
However, as the statistical uncertainties of these measurements improve, the careful analysis of the
consistency of the results derived from different data sets 
becomes crucial as it can be used to detect the presence of systematic errors.
Here we review the constraints on the parameters of the $\Lambda$CDM model obtained by combining 
our BOSS clustering measurements with different data sets.

The blue dashed lines in Fig.~\ref{fig:om_h} correspond to the constraints 
in the $\Omega_{\rm m}-h$ plane derived from the Planck (upper panel) and WMAP9 (lower panel) CMB measurements. 
Both sets of constraints are elongated along the same degeneracy, which approximately corresponds to 
a constant value of $\Omega_{\rm m}h^3$ \citep{Percival2002,PlanckXVI2013}, indicated by
the dotted lines. Along this degeneracy, the constraints from WMAP9 extend towards lower values of
$\Omega_{\rm m}$ and higher values of $h$ than those derived from the Planck data. This behaviour leads to different,
but consistent, marginalized constraints on these parameters. 
As shown by the red solid lines in Fig.~\ref{fig:om_h}, when these CMB measurements are combined with the
information from the LOWZ and CMASS DR11 clustering wedges the obtained constraints are significantly improved,
leading to  $\Omega_{\rm m}=0.283\pm0.010$ and $h=0.6947\pm0.0097$ for the WMAP9+BOSS $\xi_{\Delta \mu}(s)$
combination and $\Omega_{\rm m}=0.2974\pm0.0098$ and $h=0.6859\pm0.0076$ for the
Planck+BOSS $\xi_{\Delta \mu}(s)$ data set. Although the differences in 
the constraints from Planck and WMAP9 are propagated to these results, our BOSS DR11 clustering measurements
select the lowest values of $\Omega_{\rm m}$ allowed by Planck, leading to final constraints 
which are consistent within one~$\sigma$ with those derived using WMAP9. 
As shown in Table~\ref{tab:extra}, using the angle-averaged correlation function of the LOWZ and CMASS DR11 
samples leads to similar constraints. 

The Dashed lines in Figs.~\ref{fig:measurdr10} and \ref{fig:measurdr11} correspond to the best-fitting
$\Lambda$CDM model to the Planck+BOSS $\xi_{\Delta \mu}(s)$ combination. This model provides an excellent
description of the broad band shape and the location of the BAO peak in these measurements, with $\chi^2$ values of 
49.7 and 48.3 over 52 degrees of freedom for the DR11 LOWZ and CMASS clustering wedges, respectively.
Despite the fact that this model was obtained by fitting the clustering wedges of the DR11 galaxy samples
it also provides an excellent description of all our clustering measurements, including our DR10 results.
This illustrates the consistency between these data sets, which can also be seen in the cosmological
constraints obtained when the Planck CMB data is combined with the DR10 LOWZ and CMASS clustering wedges,
in which case we find $\Omega_{\rm m}=0.294\pm0.010$ and $h=0.6882\pm0.0079$, in good agreement with the results obtained using DR11 information. 

Our tightest constraints on the parameters of the $\Lambda$CDM model are obtained by combining the
ePlanck CMB data set with the information from our DR11 BOSS $\xi_{\Delta \mu}(s)$, SN and
BAO data sets, leading to $\Omega_{\rm m}=0.2924\pm0.0086$ and $h=0.6899\pm0.0070$

\subsection{The dark energy equation of state}
\label{sec:wde}

In the standard $\Lambda$CDM model, the current phase of accelerated cosmic expansion is due to a 
dark energy component characterized by a constant equation of state $w_{\rm DE}=-1$.
As this hypothesis is consistent with all current cosmological observations, it has become the standard 
model for dark energy. However, a variety of alternative models have been proposed
\citep[for a review see e.g., ][]{Peebles2003,Frieman2008}. Here we explore the
constraints on more general dark energy models by allowing for variations in $w_{\rm DE}$ and its
possible evolution with time.

We start our analysis by extending the $\Lambda$CDM parameter space including $w_{\rm DE}$,
assumed constant in time, as a free parameter. 
The blue long-dashed contours in Fig.~\ref{fig:wde} correspond to the 68 and 95 per cent confidence
levels in the $\Omega_{\rm m}$--$w_{\rm DE}$ plane obtained in this case from the WMAP9 (right panel) and 
Planck (left panel) CMB data. The constraints derived from both of these data sets exhibit a 
degeneracy between these parameters. However, the strong degeneracy seen in the WMAP9 constraints is
somewhat reduced in the results derived from Planck, as the information from the higher acoustic peaks
restricts the region of the parameter space with $w_{\rm DE}>-1$.
When these CMB data sets are combined with the information from the DR11 LOWZ and CMASS angle-averaged correlation
functions, the allowed region for these parameters is reduced to a narrow degeneracy that is mostly driven by
the $D_{\rm V}/r_{\rm d}$ constraint provided by the CMASS sample (grey long-dashed lines in Fig.~\ref{fig:wde}).
In these cases the dark energy equation of state is only weakly constrained, with $w_{\rm DE}=-1.05_{-0.13}^{+0.29}$ 
and $w_{\rm DE}=-1.28_{-0.16}^{+0.24}$ for the WMAP9+BOSS $\xi(s)$ and Planck+BOSS $\xi(s)$ combinations, 
respectively.

\begin{figure}
\includegraphics[width=0.45\textwidth]{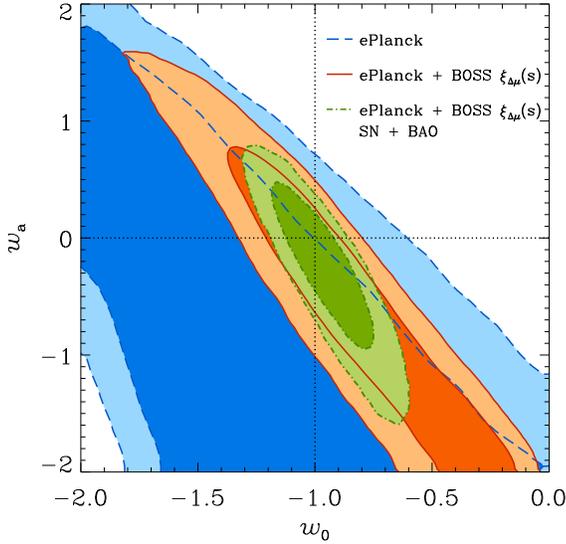}
\caption{
Marginalized 68 and 95 per cent CL in the $w_0$--$w_a$ plane when  
we explore the redshift dependence of the dark energy equation of state, 
parametrized as in equation~(\ref{eq:wa}). The contours show the results obtained using the
ePlanck CMB data alone (blue dashed lines), the ePlanck+BOSS $\xi_{\Delta \mu}(s)$ combination 
(red solid lines), and when this information is combined with our BAO and SN data sets (green dot-dashed lines).
The dotted lines correspond to the fiducial values of these parameters in the $\Lambda$CDM model,
$w_0=-1$ and $w_a=0$.
}
\label{fig:wa} 
\end{figure}

The red solid lines in Fig.~\ref{fig:wde} show the constraints obtained when the WMAP9 and Planck CMB data sets are combined with the
information from the full shape of the DR11 LOWZ and CMASS clustering wedges. The additional information 
provided by $\xi_{\perp}(s)$ and $\xi_{\parallel}(s)$ can break the degeneracy present in the CMB results much more efficiently than
the angle-averaged correlation function, leading to broadly similar results for both CMB data sets.
In particular, the marginalized constraints on the matter density parameter are almost identical, with $\Omega_{\rm m}=0.288\pm0.015$ 
for the WMAP9+BOSS $\xi_{\Delta \mu}(s)$ case and $\Omega_{\rm m}=0.289\pm0.016$ for the Planck+BOSS $\xi_{\Delta \mu}(s)$ combination.
However, the differences in the CMB data sets lead to slightly different constraints on the dark energy equation
of state of $w_{\rm DE}=-0.964\pm0.077$ (WMAP9+BOSS $\xi_{\Delta \mu}(s)$) and $w_{\rm DE}=-1.049\pm0.078$
(Planck+BOSS $\xi_{\Delta \mu}(s)$). 
These results show that the combination of current CMB and LSS data sets can constrain the dark energy equation of state 
with an accuracy of 8 per cent, leading to results in good agreement with a cosmological constant,
indicated by the dotted line in Fig.~\ref{fig:wde}.
Our results are consistent with those reported in our companion papers \citep{Aardwolf2013,Samushia2013,Chuang2013b},
who find similar constraints on $w_{\rm DE}$ from the combination of CMB data with various types of anisotropic clustering information from the DR11 CMASS and LOWZ samples. 
This agreement illustrates the robustness of these limits with respect to the methodology implemented
to obtain them.

Although the constraints obtained using the WMAP9 and Planck data sets are consistent at the one~$\sigma$ level,
the difference between these results highlights the importance of understanding the origin of the
discrepancies between these data sets. The same behaviour is seen in other parameter spaces, once combined with
our measurements of the LOWZ and CMASS clustering wedges, the WMAP9 and Planck CMB data sets give similar results,
although the mean values are shifted by up to one~$\sigma$. In the following sections we focus on the Planck CMB measurements and derive constraints using the ePlanck combination, but we compare the results with 
those derived using WMAP9 in some particular cases.

Combining the Planck CMB data with our DR10 clustering measurements leads to consistent results.
The combination of Planck and the DR10 LOWZ and CMASS
$\xi_{\Delta \mu}(s)$ provides the constraints $\Omega_{\rm m}=0.279\pm0.019$ and $w_{\rm DE}=-1.092_{-0.088}^{+ 0.092}$.
The smaller statistical uncertainties associated with the DR11 LOWZ and CMASS 
clustering measurements lead to a reduction of $\sim15$ per cent in the constraints on the dark energy
equation of state.
As the same agreement is seen in all cosmological parameter spaces, from now on we focus on the results
obtained using the DR11 galaxy samples. 

Including the information from the high-$\ell$ CMB experiments improves the constraints only marginally. 
Using the ePlanck + BOSS $\xi_{\Delta \mu}(s)$ combination we find $w_{\rm DE}=-1.051\pm0.076$.
Our final constraints are obtained when the information from the additional
BAO and Union2.1 SN measurements are added to this data combination, leading to 
$w_{\rm DE}=-1.024\pm0.052$, in good agreement with the 
$\Lambda$CDM model value of $w_{\rm DE}=-1$, and $\Omega_{\rm m}=0.293\pm0.011$. 
Replacing the information from the Union2.1 SN sample by the SNLS data leads to a change in the recovered
values of about one~$\sigma$, with $w_{\rm DE}=-1.071\pm0.055$ and $\Omega_{\rm m}=0.283\pm0.011$, showing 
a preference for values of $w_{\rm DE}<-1$. As pointed out by \citet{PlanckXVI2013}, the difference
between the results obtained using these samples might indicate that the treatment of the systematic
errors affecting these SN data sets is incomplete. 

Although exploring the constraints on a constant $w_{\rm DE}$ could indicate a deviation from the standard
$\Lambda$CDM paradigm, more general dark energy models, such as those based on a scalar field, will be
characterized by a time-dependent equation of state \citep[e.g.][]{Wetterich1988}.
We explore the constraints on the time-dependence of $w_{\rm DE}$, parametrized as in equation~(\ref{eq:wa}).
The blue dashed lines in Fig.~\ref{fig:wa} correspond to the two-dimensional marginalized constraints 
in the $w_0$--$w_a$ plane obtained from the ePlanck CMB, covering a large region of the parameter space. 
The red solid lines in Fig.~\ref{fig:wa}
correspond to the results obtained by combining the ePlanck CMB measurements with our BOSS $\xi_{\Delta \mu}(s)$ data set,
showing a significant reduction of the allowed region for these parameters.
In this case we find $w_0=-0.83_{-0.34}^{+0.38}$ and $w_a=-0.61_{-0.96}^{+0.89}$.
As shown by the green dot-dashed lines in the same figure, the information from our Full data set combination 
tightens the constraints, leading to $w_0=-0.95\pm0.14$ and $w_a=-0.29\pm0.47$, in agreement with the standard
$\Lambda$CDM model values indicated by the dotted lines.
 
\subsection{Non-flat universes}
\label{sec:omk}

\begin{figure}
\includegraphics[width=0.45\textwidth]{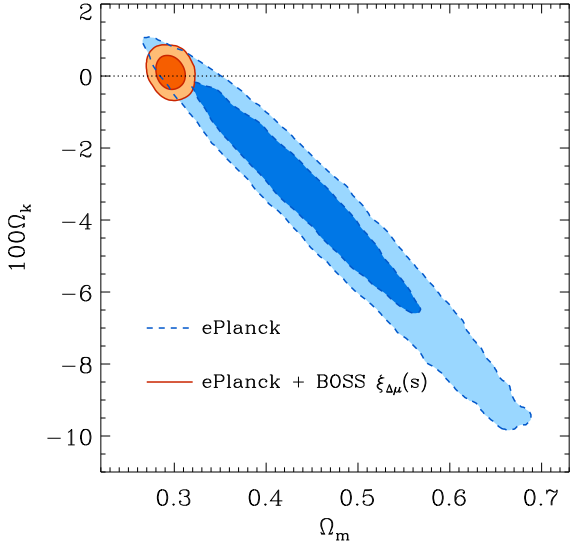}
\caption{
Marginalized constraints in the $\Omega_{\rm m}$--$\Omega_k$ plane when  
the $\Lambda$CDM model is extended to allow for non-flat models. The contours show the
68 and 95 per cent CL obtained using the ePlanck CMB data alone (blue dashed lines)
and the ePlanck+BOSS $\xi_{\Delta \mu}(s)$ combination (red solid lines).
The dotted line corresponds to flat universes, with $\Omega_k=0$.}
\label{fig:omk} 
\end{figure}

The standard $\Lambda$CDM model assumes a flat universe. Here we test this assumption by adding 
$\Omega_k$ to the list of free parameters of our base model. The blue dashed contours in
Fig.~\ref{fig:omk} show the 68 and 95 per cent marginalized constraints in the $\Omega_{\rm m}$--$\,\Omega_k$
plane derived by means of the ePlanck CMB data combination, which exhibit the so-called geometrical degeneracy
\citep{Efstathiou1999} relating models with the same angular scale of
the acoustic peaks in the CMB.
This degeneracy extends over a wide range of values of $\Omega_k$,
leading to weak constraints on this parameter. In this case we find $100\Omega_k=-4.2_{-1.7}^{+2.7}$. 
As shown by \citet{PlanckXVI2013},
including information from the lensing signal inferred from the CMB partially reduces this degeneracy,
but to obtain significantly tighter constraints it is necessary to combine these measurements with
additional data sets.

\begin{figure}
\includegraphics[width=0.45\textwidth]{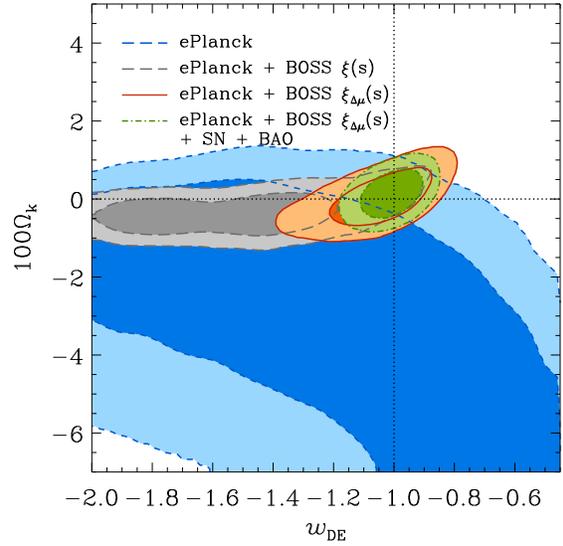}
\caption{
Marginalized constraints in the $w_{\rm DE}$--$\Omega_k$ plane for the $\Lambda$CDM 
parameter set extended by allowing for simultaneous variations on both of these parameters.
The contours correspond to the 68 and 95 per cent CL derived from the combination of ePlanck data alone
(blue dashed lines), ePlanck plus the clustering wedges of the LOWZ ans CMASS DR11 samples (red solid lines),
and when the BAO and SN data sets are added to the later combination (green dot-dashed lines). 
The dotted lines correspond to the values of these parameters in the $\Lambda$CDM model.
}
\label{fig:wok} 
\end{figure}

When the ePlanck CMB data are combined with the monopole correlation functions of the DR11
LOWZ and CMASS samples, the constraints on $D_{\rm V}/r_{\rm d}$ provided by these measurements are
sufficient to break the geometrical degeneracy, leading to a constraint of $100\Omega_k=0.07\pm0.31$,
in excellent agreement with a flat Universe.  
The red solid lines in Fig.~\ref{fig:omk} show the constraints obtained when the ePlanck data are combined
with the clustering wedges of these galaxy samples, leading to a similar result of $100\Omega_k=0.10\pm0.29$.
Including our SN and BAO data sets leads only to a small shift in the recovered mean value for this parameter,
with $100\Omega_k=0.15\pm0.29$, showing no evidence for a deviation from the flat Universe hypothesis, 
indicated by a dotted line in Fig.~\ref{fig:omk}.

Both $\Omega_k$ and $w_{\rm DE}$ are involved in the geometrical degeneracy, as they change the 
distance to the last-scattering surface. This means that when both of these parameters are varied simultaneously,
the geometric degeneracy gains an extra degree of freedom, leading to a significant degradation of the 
obtained constraints. This effect is shown by the blue short-dashed contours in Fig.~\ref{fig:wok},
which correspond to the constraints in the $w_{\rm DE}$\,--\,$\Omega_k$ plane obtained from the ePlanck CMB data. 
In this case, the information from the LOWZ and CMASS angle-averaged correlation functions is not enough to 
break the geometric degeneracy completely (as shown by the grey long-dashed lines in Fig.~\ref{fig:wok}).
Although this information can constrain the curvature of the Universe to $100\Omega_k=-0.38_{-0.28}^{+0.24}$,
it leaves a wide range
of allowed values for the dark energy equation of state in the region where $w_{\rm DE} < -1$.

The red solid contours in Fig.~\ref{fig:wok} correspond to constraints obtained after combining the
ePlanck CMB data with the full shape of the clustering wedges of the DR11 LOWZ and CMASS samples.
The additional information in the clustering wedges reduces the allowed region of this parameter
space significantly, leading to the constraints $100\Omega_k=0.02\pm0.43$ and 
 $w_{\rm DE}=-1.05\pm0.11$. 
As shown by the green dot-dashed lines in Fig.~\ref{fig:wok}, including the information from the SN and
additional BAO measurements can improve the constraints even further, leading
to $100\Omega_k=-0.14\pm0.33$ and $w_{\rm DE}=-1.009_{-0.060}^{+0.062}$,
in excellent agreement with the $\Lambda$CDM model.
In particular, the constraints on the dark energy equation of state obtained in this case have a similar 
accuracy as those presented in Sec.~\ref{sec:wde} under the assumption of a flat universe.

\begin{figure}
\includegraphics[width=0.45\textwidth]{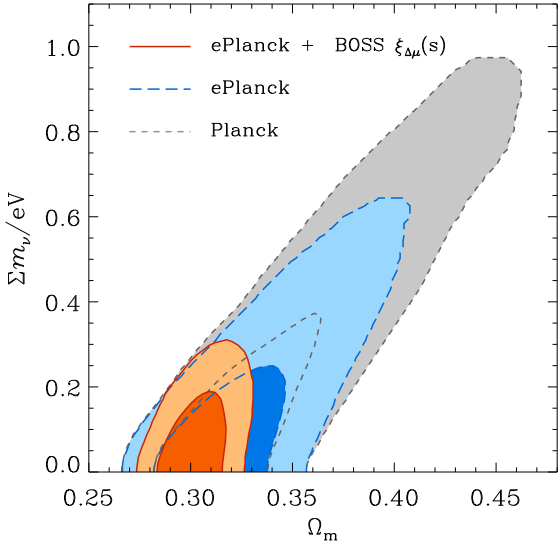}
\caption{
Marginalized constraints in the $\Omega_{\rm m}$--$\Sigma m_{\nu}$ plane
obtained when $\Lambda$CDM parameter set is extended by treating the neutrino mass as a free parameter.
The short- and lon-dashed lines correspond to the 68 and 95 per cent CL derived
by the Planck and ePlanck CMB data, respectively. The solid lines show the results obtained from the 
combination of the ePlanck CMB measurements with the full shape of the LOWZ and
CMASS clustering wedges (red solid lines).
}
\label{fig:mnu} 
\end{figure}

\subsection{Massive neutrinos and relativistic species}
\label{sec:fnu}

In recent years, neutrino oscillation experiments have measured non-zero
mass-squared differences between neutrino flavours, implying that they are massive and contribute
to the total energy budget of the Universe. However, absolute neutrino mass measurements are more
difficult to perform in the laboratory \citep{Lobashev2003,Eitel2005,Otten2008}
and current constraints are weaker than those imposed by cosmological observations such as
CMB and LSS measurements \citep{Lesgourgues2012}.

Following the analysis of \citet{PlanckXVI2013}, our base $\Lambda$CDM model includes a non-zero
contribution from massive neutrinos with $\sum m_{\nu} = 0.06\,{\rm eV}$ to the total energy budget of
the Universe. In this section we allow this parameter to vary freely assuming three neutrino species of equal
mass and explore the constraints that can be imposed on this quantity by means of our DR11 clustering
measurements.

\begin{figure}
\includegraphics[width=0.45\textwidth]{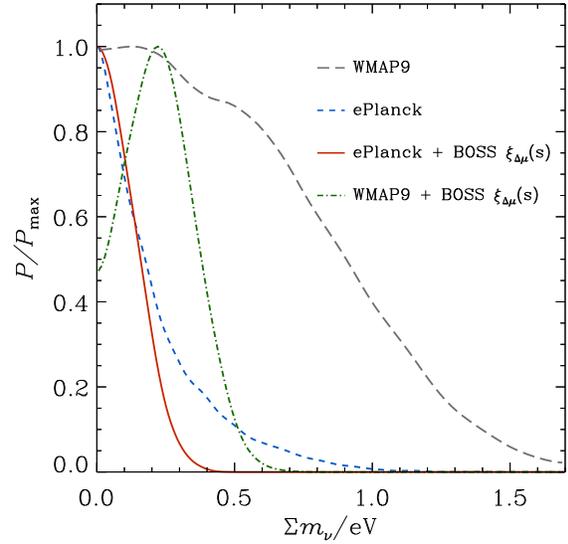}
\caption{
One-dimensional marginalized constraints on $\sum m_{\nu}$ obtained by means of the ePlanck CMB measurements
(blue short-dashed line) and the combination of these data with the BOSS $\xi_{\Delta \mu}(s)$ (red solid line).
The grey long-dashed and green dot-dashed lines correspond to the results obtained when the ePlanck
CMB measurements are replaced by WMAP9. 
}
\label{fig:mnu1d} 
\end{figure}

The grey short-dashed lines in Fig.~\ref{fig:mnu} correspond to the 68 and 95 per cent CL in the 
$\Omega_{\rm m}$--$\sum m_{\nu}$ plane derived from the Planck CMB data. These constraints are elongated along
a line that corresponds to models with a constant redshift of matter-radiation equality, $z_{\rm eq}$, which 
is accurately measured from CMB observations \citep{Komatsu2009}. As indicated by the blue long-dashed lines 
in the same figure, extending these data with the high-$\ell$ CMB measurements from ACT and SPT improves the
results significantly, but leaves a residual degeneracy that limits the constraints on $\sum m_{\nu}$. This
is shown by the blue short-dashed line in Fig.~\ref{fig:mnu1d}, which corresponds to the one-dimensional marginalized
constraints on $\sum m_{\nu}$ obtained from the ePlanck CMB data, corresponding to 
$\sum m_{\nu}< 0.66 $ eV (95 per cent CL).

The constraints on the total neutrino mass can be improved by combining the CMB information with our isotropic and
anisotropic galaxy clustering measurements. As this information improves the constraints on $\Omega_{\rm m}$, 
it can break the degeneracy present in the CMB-only constraints. This effect is illustrated by the red solid lines 
in Fig.~\ref{fig:mnu}, which correspond to the results derived from our ePlanck + BOSS $\xi_{\Delta \mu}(s)$ combination.
The effect of the extra information from BOSS on the constraints on the neutrino mass is shown by the  
red solid line in Fig.~\ref{fig:mnu1d}. In this case we obtain the limit
$\sum m_{\nu}< 0.24 $ eV (95 per cent CL). As shown in Table~\ref{tab:extra}, 
combining the CMB information with the LOWZ and CMASS angle-averaged correlation functions leads to a similar
constraint. This limit is not improved by including the additional BAO and Union2.1 SN information in
the analysis, in which case we obtain $\sum m_{\nu}< 0.23 $ eV (95 per cent CL). 
When the SNLS SN compilation is used instead of the Union2.1 sample we find a slightly
tighter constrain, with $\sum m_{\nu}< 0.21 $ eV (95 per cent CL). 

The grey long-dashed lines in Fig.~\ref{fig:mnu1d} correspond to the marginalized constraints on 
$\Sigma m_{\nu}$ obtained using WMAP9 data alone, which extend over a wide range of allowed values.
The green dot-dashed line shows the result obtained when the WMAP9 data is combined with the
LOWZ and CMASS DR11 clustering wedges. Interestingly, in this case the constraints show a preference
for a non-zero value of $\Sigma m_{\nu}\simeq0.2$ eV. This is due to a slight difference in the 
values of $z_{\rm eq}$ preferred by the Planck and WMAP9 data. When using WMAP9 data, the 
degeneracy seen in Fig.~\ref{fig:mnu} is shifted towards lower values of $\Omega_{\rm m}$. 
Adding the information from the clustering wedges measured from BOSS helps to tighten 
the constraints on the matter density, breaking the degeneracy obtained
from the CMB at a region that shows a slight preference for non-zero values of $\Sigma m_{\nu}$. 

\begin{figure}
\includegraphics[width=0.45\textwidth]{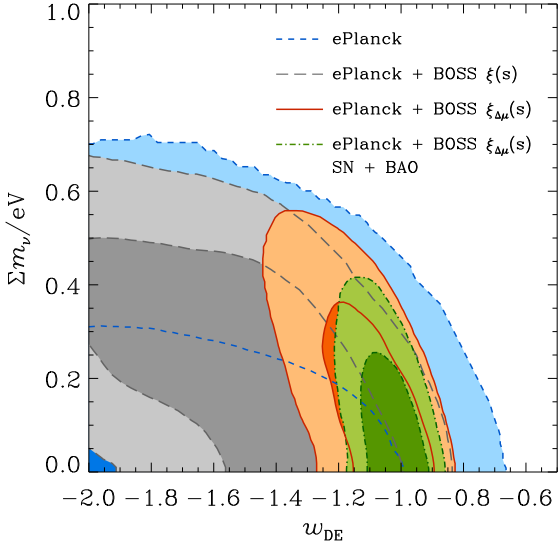}
\caption{
Marginalized constraints in the $w_{\rm DE}$--$\Sigma m_{\nu}$ plane
obtained when $\Lambda$CDM parameter set is extended by treating these quantities as free parameters.
The contours correspond to the 68 and 95 per cent CL derived by the ePlanck CMB data alone (blue short-dashed lines)
and the ePlanck+BOSS $\xi(s)$ (grey long-dashed lines),
ePlanck+BOSS $\xi_{\Delta \mu}(s)$ (red solid lines), and Full data combinations (green dot-dashed lines).
}
\label{fig:wmnu} 
\end{figure}

The constraints on the total neutrino mass derived from cosmological observations are model dependent,
as they vary depending on the parameter space being studied \citep{Zhao2013}. For example, if the dark
energy equation of
state is allowed to vary, the degeneracy in the CMB constraints gains an extra degree of freedom. This
behaviour can be seen in the blue dashed contours of Fig.~\ref{fig:wmnu}, which correspond to the
two-dimensional marginalized constraints in the $w_{\rm DE}$--$\sum m_{\nu}$ plane derived 
from the ePlanck data set. As shown by the grey long-dashed lines, in this parameter space the information
provided by our BOSS $\xi(s)$ measurements cannot break the CMB degeneracy efficiently, 
leading to poor marginalized constraints. 
The additional information in the full shape of the clustering wedges reduces the allowed region
for these parameters significantly, leading to the limit $\sum m_{\nu}< 0.47 $ eV (95 per cent CL)
and $w_{\rm DE}=-1.13\pm0.12$. 
After also including the BAO and SN information the constraints on the neutrino mass are improved to
$\sum m_{\nu}< 0.33 $ eV (95 per cent CL). This demonstrates that, when the uncertainties in the exact value
of $w_{\rm DE}$ are taken into account, the upper bound on $\sum m_{\nu}$ is increased by 50 per cent with 
respect to the one obtained under the assumption of a $\Lambda$CDM model. 

\begin{figure}
\includegraphics[width=0.45\textwidth]{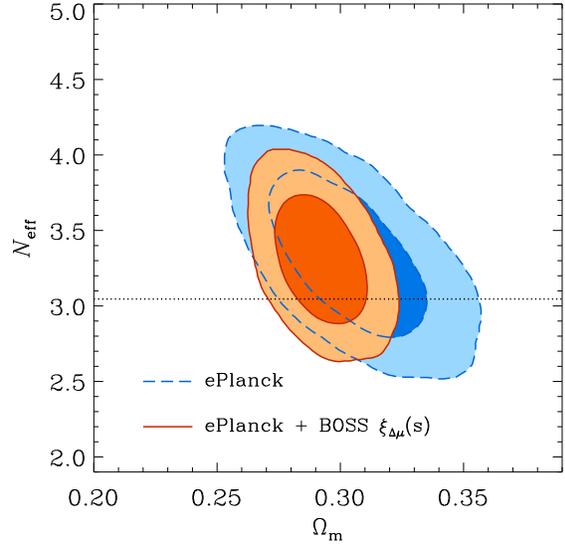}
\caption{
Marginalized constraints in the $\Omega_{\rm m}$--$N_{\rm eff}$ plane
obtained when $\Lambda$CDM parameter set is extended by treating the effective number of relativistic 
species as a free parameter.
The dashed and solid lines correspond to the 68 and 95 per cent CL derived
by the ePlanck CMB data and its combination with the full shape of the DR11 LOWZ and
CMASS clustering wedges. The dotted line indicates the standard value of $N_{\rm eff}=3.046$ 
}
\label{fig:nnu} 
\end{figure}

It is also interesting to explore for potential deviations on the effective number of relativistic species
from its standard value $N_{\rm eff}=3.046$. For this analysis we extend the $\Lambda$CDM parameter space including 
$N_{\rm eff}$ as a free parameter, while assuming that the additional relativistic species are massless. 
As discussed in \cite{PlanckXVI2013}, the ePlanck CMB combination can place tight constraints on this parameter.
These constraints are shown by the blue dashed lines in Fig.~\ref{fig:nnu} which correspond to our ePlanck constraints
in the $\Omega_{\rm m}$--$N_{\rm eff}$ plane, leading to $N_{\rm eff}=3.35\pm0.33$. Adding the 
information from the LOWZ and CMASS angle-averaged correlation functions leads to
an improvement of this limit, with $N_{\rm eff}=3.31\pm0.27$. Using the information from our BOSS
$\xi_{\Delta \mu}(s)$ or adding the information from the BAO and SN data leaves this result essentially unchanged.

\subsection{Constraining deviations from general relativity}
\label{sec:fg}

In Sections~\ref{sec:lcdm}--\ref{sec:fnu} we computed the logarithmic
growth $f(z)$ required for our model of the full shape of the clustering wedges in the context of GR.
Here we relax this assumption and parametrize its redshift evolution as 
$f(z)=\Omega_{\rm m}(z)^{\gamma}$, with the exponent $\gamma$ treated as a free parameter.
As described in \citet{Linder2007}, GR predicts a value of $\gamma\simeq0.55$, with small corrections
depending on the value of $w_{\rm DE}$. Then, the constraints on this parameter
from the full shape of the LOWZ and CMASS clustering wedges can be used to set limits on potential
deviations from the predictions of GR \citep{Guzzo2008}.
This analysis is only possible using anisotropic clustering measurements such as the clustering wedges,
as the effect of $f(z)$ and the bias parameter are degenerate in angle-averaged quantities.

\begin{figure} 
\includegraphics[width=0.45\textwidth]{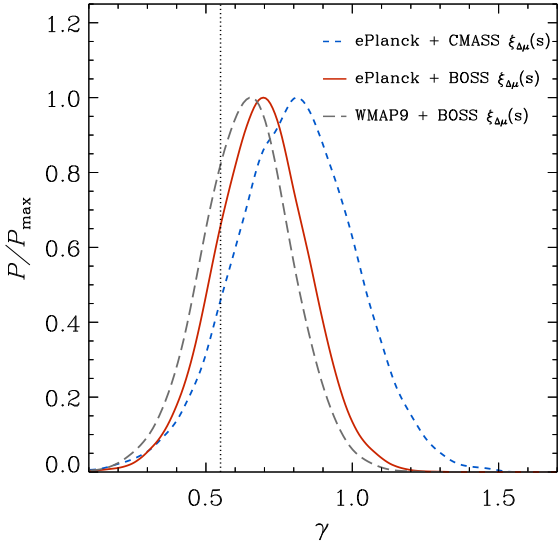}
\caption{
One-dimensional marginalized constraints on the power-law index of the structure growth rate parameter, 
assuming $f(z)=\Omega_{\rm m}^\gamma$.
The blue dashed lines correspond to the results obtained by combining the ePlanck CMB
data with the DR11 CMASS clustering wedges. The red solid line shows the improvement obtained
by including the information of the clustering wedges of the LOWZ sample in the analysis.
Replacing the ePlanck CMB measurements by the WMAP9 data leads to similar constraints (grey dashed lines).
}
\label{fig:fg} 
\end{figure}

The blue short-dashed line in Fig.~\ref{fig:fg} corresponds to the one-dimensional marginalized constraints on
$\gamma$ obtained from the combination of the Planck CMB measurements with the DR11 CMASS $\xi_{\Delta}(s)$.
In this case we obtain $\gamma=0.77\pm0.20$. Although a wide range of values of this
parameter are allowed by the data, these results are consistent within one\,$\sigma$ with the predictions 
from GR, indicated by the dotted line. 
As shown by the red solid line, adding the information from the clustering wedges of the LOWZ sample 
improves the constraints to $\gamma=0.69\pm0.15$, which illustrates the importance of including the 
low-redshift measurements. As shown in Table~\ref{tab:extra}, these constraints are not modified
when the Planck data are extended with the high-$\ell$ CMB measurements, or when the BAO and SN data sets
are included in the analysis.
The grey long-dashed line in Fig.~\ref{fig:fg} shows the constraints obtained by means of the
WMAP9+BOSS $\xi_{\Delta}(s)$ combination, which leads to a constraint of $\gamma=0.64\pm0.15$,
in good agreement with the result obtained using the data from the Planck satellite.

Our constraints are in good agreement with those inferred in our companion papers. 
\citet{Samushia2013b} use the full shape of the CMASS monopole-quadrupole pair in combination with 
recent CMB measurements to find $\gamma=0.69\pm0.11$, while \citet{Beutler2013} derive a constraint
of $\gamma=0.772_{-0.097}^{+0.124}$ from the combination of Planck data with the multipoles of the CMASS power 
spectrum. 

\begin{figure}
\includegraphics[width=0.45\textwidth]{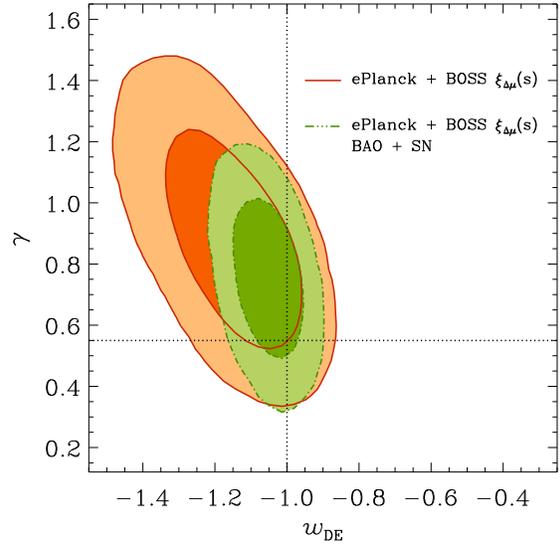}
\caption{
Marginalized constraints in the $w_{\rm DE}$--$\gamma$ plane
obtained when $\Lambda$CDM parameter set is extended by treating these quantities as free parameters.
The red solid lines correspond to the 68 and 95 per cent CL obtained by combining the ePlanck CMB
data with the DR11 LOWZ and CMASS clustering wedges. The green dot-dashed
lines show the result of including the additional BAO and SN data sets in the analysis.
}
\label{fig:fgw} 
\end{figure}

We also tested the effect of allowing for simultaneous variations
of $w_{\rm DE}$ (assumed time independent) and $\gamma$. Fig.~\ref{fig:fgw} presents the two-dimensional
marginalized constraints in the $w_{\rm DE}$--$\gamma$ plane obtained in this case
by means of the ePlanck + BOSS $\xi_{\Delta \mu}(s)$ combination (solid lines), and when these data are combined
with the BAO and SN data sets (dot-dashed lines). 
Including $\gamma$ as a free parameter leads to a degeneracy
between this quantity and the dark energy equation of state, degrading the constraints on these parameters.
In this case we find $w_{\rm DE}=-1.15\pm0.11$ and $\gamma=0.88\pm0.22$. 
As discussed in \citet{Sanchez2013}, assuming that $f(z_{\rm m})$ follows the predictions of GR implies that 
the relative amplitude of $\xi_{\perp}(s)$ and $\xi_{\parallel}(s)$ provides information on $\Omega_{\rm m}$ 
that improves the obtained constraints. However, when this assumption is relaxed including 
$\gamma$ as a free parameter, this extra constraining powers is lost, leading to weaker limits.
Including the SN and additional BAO measurements reduces the degeneracy present in the
CMB + BOSS $\xi_{\Delta \mu}(s)$ constraints,
leading to $w_{\rm DE}=-1.055\pm0.057$  and $\gamma=0.75\pm0.17$,
in agreement with the results obtained when these parameters are varied separately.

\section{Conclusions}
\label{sec:conclusions}

We have analysed the cosmological implications of the angle-averaged correlation functions, $\xi(s)$,
and the clustering wedges, $\xi_{\perp}(s)$ and $\xi_{\parallel}(s)$, of the LOWZ and CMASS
samples corresponding to SDSS-DR10 and DR11.
We use a simple parametrization, based on renormalized perturbation theory, as a tool to extract cosmological
information from the full shape of these measurements for $s \gtrsim 40\,\mpc$.
We combine this information with CMB, SN, and additional BAO measurements to derive constraints on
the parameters of the standard $\Lambda$CDM model and a number of potential extensions, including curvature,
alternative dark energy models, massive neutrinos, additional relativistic species and deviations from the
predictions of general relativity.
As shown by \citet{Sanchez2013}, we find that the extra information provided by the clustering wedges
is most useful when the dark energy equation of state is treated as a free parameter.

The constraints on $H(z)$ and $D_{\rm A}(z)$ from the clustering wedges are consistent 
with the predictions of the best-fitting $\Lambda$CDM model to the Planck CMB measurements.
Assuming that the growth of structures follows the predictions of GR, the full
shape of the LOWZ clustering wedges imply 
$D_{\rm A}(z_{\rm m})= (965\pm37) \left(r_{\rm d}/r_{\rm d}^{\rm fid}\right)\,{\rm Mpc}$ and
$H(z_{\rm m})=(82.5\pm3.5)\left(r_{\rm d}^{\rm fid}/r_{\rm d}\right)\,{\rm km}\,{\rm s}^{-1}{\rm Mpc}^{-1}$
at the mean redshift $z_{\rm m}=0.32$, while the CMASS results give 
$D_{\rm A}(z_{\rm m})= (1387\pm22) \left(r_{\rm d}/r_{\rm d}^{\rm fid}\right)\,{\rm Mpc}$ and
$H(z_{\rm m})=(94.3\pm2.4)\left(r_{\rm d}^{\rm fid}/r_{\rm d}\right)\,{\rm km}\,{\rm s}^{-1}{\rm Mpc}^{-1}$
at $z_{\rm m}=0.57$.
Relaxing the assumption of GR, we
find the marginalized constraints of $f\sigma_8(z=0.32)=0.48\pm0.10$ and $f\sigma_8(z=0.57)=0.417\pm0.045$.
These values are in good agreement with those reported in our companion papers
\citep{Aardwolf2013,Chuang2013b,Samushia2013b,Beutler2013},
indicating the robustness of our results with respect to details in the methodology implemented in the analysis. 

As can be seen in Table~\ref{tab:extra},
our results show no significant evidence for a deviation from the standard $\Lambda$CDM model, 
which provides a good description of the full shape of all our clustering measurements.
In particular, the ePlanck+BOSS $\xi_{\Delta \mu}(s)$ combination alone is sufficient to constrain the curvature of
the Universe to $\Omega_k =0.0010\pm0.0029$, the total neutrino mass to  
$\sum m_\nu< 0.24\,{\rm eV} $ (95 per cent CL), the effective number of relativistic species to 
$N_{\rm eff}=3.31\pm0.27$, and the dark energy equation of state to $w_{\rm DE}=-1.051\pm0.076$.
Adding the information from our BAO and SN data sets further improves these constraints. 

The assumption that the dark energy component can be characterized by a constant equation of state specified 
by $w_{\rm DE}(z)=-1$ has strong implications on the constraints obtained in several parameter 
spaces. Allowing this parameter to vary weakens the constraints obtained from the combination of
CMB information with our BOSS $\xi(s)$ measurements.
These cases illustrate the extra constraining power of the clustering
wedges, which lead in general to similar constraints than the ones derived under the assumption that dark energy
behaves as a cosmological constant.

The information from the full shape of the clustering wedges can be used to constrain potential deviations from 
the predictions of GR. Assuming that $f(z)=\Omega_{\rm m}(z)^\gamma$, the combination of
the ePlanck CMB measurements with the DR11 LOWZ and CMASS clustering wedges gives a
constraint of $\gamma=0.69\pm0.15$, consistent with no deviation from the GR prediction of $\gamma=0.55$
within one~$\sigma$.
The assumption that $f(z)$ follows the predictions of GR implies that 
the relative amplitude of the clustering wedges contains information on $\Omega_{\rm m}$.
When this assumption is relaxed, this additional constraining power is lost,
affecting the constraints on other parameters.
For example, if $\gamma$ and $w_{\rm DE}$ are varied simultaneously, 
the ePlanck+BOSS $\xi_{\Delta \mu}(s)$ combination implies that
$\gamma=0.88\pm0.22$ and  $w_{\rm DE}=-1.15\pm0.11$. 
When the additional BAO and Union2.1 SN measurements are included in the analysis 
we find $w_{\rm DE}=-1.055\pm0.057$ and $\gamma=0.75\pm0.17$, 
in agreement with the results obtained when these parameters are varied separately.

Combining our clustering measurements with the WMAP9 or Planck CMB measurements leads to 
results that, although consistent, differ at the one~$\sigma$ level.
This highlights the importance of a detailed analysis 
of the origin of the differences between WMAP9 and Planck as these could indicate the presence of systematic errors.
Similar differences are observed when the Union2.1 SN compilation is replaced by the SNLS sample, which
leads to changes in the obtained constraints of the same order or larger than the associated 
statistical errors. This situation might change in the near future, as results based on new calibrations of
these SN samples \citep{Betoule2013} become available. 

As SDSS-III approaches the end of observations in June 2014, the galaxy and quasar samples from BOSS are close to
being completed. Besides the improvement in the statistical power of the final data sets, these samples will
allow us to perform an improved analysis of the potential systematic errors affecting our measurements.
The final galaxy samples from BOSS will deliver the most accurate views of the LSS of the Universe at $z<0.7$,  
thus providing invaluable cosmological information.

\section*{Acknowledgements}

AGS would like to thank Ximena Mazzalay for useful discussions.
AGS and FM acknowledge support from the Trans-regional Collaborative Research 
Centre TR33 `The Dark Universe' of the German Research Foundation (DFG).
EK is supported by the Australian Research Council Centre of Excellence for
All-sky Astrophysics (CAASTRO), through project number CE110001020.

Numerical computations for the PTHalos mocks were done on the Sciama
High Performance Compute (HPC) cluster which is supported by the ICG,
SEPNet and the University of Portsmouth

Funding for SDSS-III has been provided by the Alfred P. Sloan Foundation, the Participating
Institutions, the National Science Foundation, and the U.S. Department of Energy. 

SDSS-III is managed by the Astrophysical Research Consortium for the
Participating Institutions of the SDSS-III Collaboration including the
University of Arizona,
the Brazilian Participation Group,
Brookhaven National Laboratory,
University of Cambridge,
Carnegie Mellon University,
University of Florida,
the French Participation Group,
the German Participation Group,
Harvard University,
the Instituto de Astrofisica de Canarias,
the Michigan State/Notre Dame/JINA Participation Group,
Johns Hopkins University,
Lawrence Berkeley National Laboratory,
Max Planck Institute for Astrophysics,
Max Planck Institute for Extraterrestrial Physics,
New Mexico State University,
New York University,
Ohio State University,
Pennsylvania State University,
University of Portsmouth,
Princeton University,
the Spanish Participation Group,
University of Tokyo,
University of Utah,
Vanderbilt University,
University of Virginia,
University of Washington,
and Yale University.

Based on observations obtained with Planck (http://www.esa.int/Planck), an ESA science
mission with instruments and contributions directly funded by ESA Member States, NASA, and Canada.
We acknowledge the use of the Legacy Archive for Microwave Background 
Data Analysis (LAMBDA). Support for LAMBDA is provided by the NASA
Office of Space Science. 





\appendix

\section{Summary of the obtained constraints} 
\label{sec:tables}

In this appendix we summarize the constraints on cosmological parameters obtained using different combinations of
the data sets described in Section~\ref{sec:data}.
Tables~\ref{tab:lcdm}-\ref{tab:fg} list the 68\% confidence limits obtained in the parameter spaces
analysed in Sections~\ref{sec:lcdm} to \ref{sec:fg}. The upper section of these tables lists
the constraints on the main parameters included in the fits, while the lower section contains the results on the parameters derived from the first set.


\begin{table*}  
\centering
  \caption{
    Marginalized 68\% constraints on the cosmological parameters of the standard $\Lambda$CDM model,
obtained using different combinations of the data sets described in Section~\ref{sec:data}. 
}
    \begin{tabular}{@{}lccc@{}}
    \hline
 &\multirow{2}{*}{ePlanck+BOSS $\xi(s)$}  & \multirow{2}{*}{ePlanck+BOSS $\xi_{\Delta \mu}(s)$} & ePlanck + BOSS $\xi_{\Delta \mu}(s)$  \\
                                         &                       &                             &    +BAO+SN         \\  
\hline
\multicolumn{3}{l}{Main parameters} &  \\[0.3mm]
      $100\,\omega_{\rm b}$  & $2.2202\pm0.024$        &  $2.226\pm0.025$           &  $2.227\pm0.024$    \\[0.3mm]
       $100\,\omega_{\rm c}$ & $11.78\pm0.16$          &  $11.69\pm0.17$            &  $11.66\pm0.16$      \\[0.3mm]
       $100\,\Theta$         & $104.16\pm0.0056$       &  $104.17\pm0.0058$         &  $104.17\pm0.0056$   \\[0.3mm]
       $n_{\rm s}$           & $0.9628\pm0.0052$       &  $0.9650\pm0.0055$         &  $0.9651\pm0.0052$   \\[0.3mm]
     $\ln(10^{10}A_{\rm s})$ & $3.089\pm0.025$         &  $3.086\pm0.027$           &  $3.087\pm0.025$   \\[0.3mm]
\multicolumn{3}{l}{Derived parameters} &  \\[0.3mm]
        $100\Omega_{\rm DE}$ & $69.78_{-0.94}^{+0.96}$ &  $70.288\pm0.99$           &  $70.47\pm0.91$   \\[0.3mm]
        $100\Omega_{\rm m}$  & $30.22_{-0.96}^{+0.94}$ &  $29.712\pm0.99$           &  $29.53\pm0.91$  \\[0.3mm]
        $\sigma_{8}$         & $0.822\pm0.011$         &  $0.818\pm0.012$           &  $0.817\pm0.011$   \\[0.3mm]
          $h$                & $0.6823_{-0.0071}^{+0.0073}$ &  $0.6862\pm0.0077$    &  $0.6876\pm0.0072$   \\[0.3mm]
      $t_{0}/{\rm Gyr}$      & $13.784\pm0.036$        &  $13.771\pm0.037$          &  $13.768\pm0.036$   \\[0.3mm]

\hline
\end{tabular}
\label{tab:lcdm}
\end{table*}

\begin{table*}  
\centering
  \caption{
    Marginalized 68\% constraints on the cosmological parameters of the $\Lambda$CDM model extended by including $w_{\rm DE}$ (assumed constant) as a free parameter, obtained using different combinations of the datasets described in Section~\ref{sec:data}.
}
    \begin{tabular}{@{}lccc@{}}
    \hline
 &\multirow{2}{*}{ePlanck+BOSS $\xi(s)$}  & \multirow{2}{*}{ePlanck+BOSS $\xi_{\Delta \mu}(s)$} & ePlanck + BOSS $\xi_{\Delta \mu}(s)$  \\
                                         &                       &                             &    +BAO+SN         \\  
\hline
\multicolumn{3}{l}{Main parameters} &  \\[0.3mm]
        $w_{\rm DE}$          & $-1.31_{-0.16}^{+0.21}$    &  $-1.051\pm0.077$           &  $-1.024\pm0.053$    \\[0.3mm]
       $100\,\omega_{\rm b}$  & $2.205\pm0.025$            &  $2.2217\pm0.025$           &  $2.224\pm0.025$     \\[0.3mm]
        $100\,\omega_{\rm c}$ & $12.06\pm0.23$             &  $11.76\pm0.21$             &  $11.72\pm0.20$      \\[0.3mm]
        $100\,\Theta$         & $104.12\pm0.059$           &  $104.16\pm0.060$           &  $104.17\pm0.058$    \\[0.3mm]
        $n_{\rm s}$           & $0.9563\pm0.0063$          &  $0.9630\pm0.0062$          &  $0.9639\pm0.0059$   \\[0.3mm]
      $\ln(10^{10}A_{\rm s})$ & $3.088\pm0.024$            &  $3.085\pm0.025$            &  $3.085\pm0.024$     \\[0.3mm]
\multicolumn{3}{l}{Derived parameters} &  \\[0.3mm]
         $100\Omega_{\rm DE}$ & $75.1_{-3.4}^{+2.6}$  &  $71.2\pm1.6$            &  $70.7\pm1.2$     \\[0.3mm]
         $100\Omega_{\rm m}$  & $24.9_{-2.6}^{+3.4}$  &  $28.8\pm1.6$            &  $29.2\pm1.2$     \\[0.3mm]
         $\sigma_{8}$         & $0.915_{-0.061}^{+0.052}$  &  $0.834\pm0.027$            &  $0.825\pm0.021$     \\[0.3mm]
           $h$                & $0.763_{-0.056}^{+0.041}$  &  $0.699_{-0.021}^{+0.020}$  &  $0.692\pm0.014$     \\[0.3mm]
       $t_{0}/{\rm Gyr}$      & $13.678_{-0.063}^{+0.068}$ &  $13.753\pm0.049$           &  $13.762\pm0.040$    \\[0.3mm]
\hline
\end{tabular}
\label{tab:wcdm}
\end{table*}

\begin{table*}  
\centering
  \caption{
    The marginalized 68\% constraints on the cosmological parameters of the $\Lambda$CDM model extended by allowing for
    variations on $w_{\rm DE}(a)$ (parametrized according to equation~\ref{eq:wa}),
    obtained using different combinations of the datasets described in Section~\ref{sec:data}. 
}
    \begin{tabular}{@{}lccc@{}}
    \hline
 &\multirow{2}{*}{ePlanck+BOSS $\xi(s)$}  & \multirow{2}{*}{ePlanck+BOSS $\xi_{\Delta \mu}(s)$} & ePlanck + BOSS $\xi_{\Delta \mu}(s)$  \\
                                         &                       &                             &    +BAO+SN         \\  
\hline
\multicolumn{3}{l}{Main parameters} &  \\[0.3mm]
     $w_0$                        & $1.29_{-0.46}^{+0.50}$ &  $-0.83_{-0.34}^{+0.39}$   &  $-0.95\pm0.14$ \\[0.3mm]
     $w_a$                        & $-0.2_{-1.1}^{+1.0}$    &  $-0.61_{-0.98}^{+0.90}$   &  $-0.29_{-0.49}^{+0.48}$ \\[0.3mm]
   $100\,\omega_{\rm b}$          & $2.206\pm0.025$         &  $2.219\pm0.025$           &  $2.221\pm0.025$ \\[0.3mm]
    $100\,\omega_{\rm c}$         & $12.04\pm0.23$          &  $11.80\pm0.20$            &  $11.76\pm0.20$ \\[0.3mm]
    $100\,\Theta$                 & $104.12_{-0.057}^{+0.059}$   &  $104.15\pm0.060$     &  $104.16\pm0.060$ \\[0.3mm]
    $n_{\rm s}$                   & $0.9565_{-0.0064}^{+0.0063}$ &  $0.9618\pm0.0062$    &  $0.9628\pm0.0062$ \\[0.3mm]
  $\ln(10^{10}A_{\rm s})$         & $3.087\pm0.024$              &  $3.084\pm0.025$      &  $3.083\pm0.025$  \\[0.3mm]
\multicolumn{3}{l}{Derived parameters} &  \\[0.3mm]
     $100\Omega_{\rm DE}$         & $74.8_{-5.3}^{+4.9}$    &  $69.1_{-4.1}^{+3.6}$   &  $70.5\pm1.3$ \\[0.3mm]
     $100\Omega_{\rm m}$          & $25.2_{-4.9}^{+5.3}$     &  $30.9_{-3.6}^{+4.1}$   &  $29.5\pm1.3$ \\[0.3mm]
     $\sigma_{8}$                 & $0.912_{-0.071}^{+0.068}$    &  $0.821_{-0.39}^{+0.38}$     &  $0.827\pm0.022$ \\[0.3mm]
       $h$                        & $0.763\pm0.079$              &  $0.679_{-0.045}^{+0.039}$   &  $0.690\pm0.014$ \\[0.3mm]
   $t_{0}/{\rm Gyr}$              & $13.683_{-0.070}^{+0.069}$   &  $13.753\pm0.054$           &  $13.748\pm0.049$ \\[0.3mm]
\hline
\end{tabular}
\label{tab:wacdm}
\end{table*}

\begin{table*}  
\centering
  \caption{
    Marginalized 68\% constraints on the cosmological parameters of the $\Lambda$CDM model extended by including non-flat models, obtained using different combinations of the datasets described in Section~\ref{sec:data}.
}
    \begin{tabular}{@{}lccc@{}}
    \hline
  &\multirow{2}{*}{ePlanck+BOSS $\xi(s)$}  & \multirow{2}{*}{ePlanck+BOSS $\xi_{\Delta \mu}(s)$} & ePlanck + BOSS $\xi_{\Delta \mu}(s)$  \\
                                         &                       &                             &    +BAO+SN         \\  
\hline
   $100\,\Omega_{k}$        & $0.07\pm0.31$      &  $0.10\pm0.29$      &  $0.15\pm0.29$   \\[0.3mm]
       $100\,\omega_{\rm b}$    & $2.218\pm0.027$    &  $2.223\pm0.027$    &  $2.222\pm0.028$   \\[0.3mm]
        $100\,\omega_{\rm c}$   & $11.83\pm0.26$     &  $11.75\pm0.25$     &  $11.77\pm0.24$   \\[0.3mm]
        $100\,\Theta$           & $104.15\pm0.060$   &  $104.16\pm0.063$   &  $104.16\pm0.063$   \\[0.3mm]
        $n_{\rm s}$             & $0.9618\pm0.0070$  &  $0.9632\pm0.0069$  &  $0.9630\pm0.0068$   \\[0.3mm]
      $\ln(10^{10}A_{\rm s})$   & $3.089\pm0.024$    &  $3.085\pm0.025$    &  $3.086\pm0.025$   \\[0.3mm]
\multicolumn{3}{l}{Derived parameters} &  \\[0.3mm]
  $100\Omega_{\rm DE}$   & $69.75\pm0.97$  &  $70.3_{-1.0}^{+0.9}$  &  $70.38_{-0.93}^{+0.92}$   \\[0.3mm]
         $100\Omega_{\rm m}$    & $30.18\pm0.96$  &  $29.6\pm1.0$          &  $29.47_{-0.91}^{+0.94}$   \\[0.3mm]
         $\sigma_{8}$           & $0.824\pm0.013$    &  $0.820\pm0.013$      &  $0.821\pm0.013$   \\[0.3mm]
           $h$                  & $0.6839_{-0.0094}^{+0.0096}$  &  $0.6890\pm0.010$  &  $0.6908\pm0.0096$   \\[0.3mm]
       $t_{0}/{\rm Gyr}$        & $13.75\pm0.12$     &  $13.73\pm0.12$       &  $13.71\pm0.12$   \\[0.3mm]
\hline
\end{tabular}
\label{tab:kcdm}
\end{table*}

\begin{table*}  
\centering
  \caption{
    The marginalized 68\% constraints on the cosmological parameters of the $\Lambda$CDM model extended by allowing
for simultaneous variations on $w_{\rm DE}$ and $\Omega_k$, obtained using different combinations of the datasets described in Section~\ref{sec:data}.
}
    \begin{tabular}{@{}lccc@{}}
    \hline
 &\multirow{2}{*}{ePlanck+BOSS $\xi(s)$}  & \multirow{2}{*}{ePlanck+BOSS $\xi_{\Delta \mu}(s)$} & ePlanck + BOSS $\xi_{\Delta \mu}(s)$  \\
                                         &                       &                             &    +BAO+SN         \\  
\hline
           $100\Omega_{k}$        & $-0.38_{-0.24}^{+0.22}$  & $0.02\pm0.43$   &  $0.14\pm0.34$     \\[0.3mm]
           $w_{\rm DE}$           & $-1.53_{-0.28}^{+0.24}$  & $-1.05\pm0.11$   &  $-1.009\pm0.063$  \\[0.3mm]
        $100\,\omega_{\rm b}$     & $2.222\pm0.028$          & $2.223\pm0.028$  &  $2.221\pm0.028$   \\[0.3mm] 
         $100\,\omega_{\rm c}$    & $11.84\pm0.25$           & $11.76\pm0.26$   &  $11.77\pm0.25$   \\[0.3mm]
         $100\,\Theta$            & $104.15\pm0.062$         & $104.16\pm0.063$ &  $104.16\pm0.063$ \\[0.3mm]
         $n_{\rm s}$              & $0.9610_{-0.0067}^{+0.0068}$  & $0.9631_{-0.0069}^{+0.0070}$   &  $0.9629\pm0.0070$  \\[0.3mm]
       $\ln(10^{10}A_{\rm s})$    & $3.085\pm0.024$            & $3.084\pm0.025$   &  $3.085\pm0.024$  \\[0.3mm]
\multicolumn{3}{l}{Derived parameters} &  \\[0.3mm] 
          $100\Omega_{\rm DE}$    & $78.4_{-3.4}^{+3.9}$  & $71.1\pm2.3$   &  $70.5\pm1.3$  \\[0.3mm]
          $100\Omega_{\rm m}$     & $21.9_{-3.9}^{+3.3}$  & $28.9\pm2.0$   &  $29.4\pm1.2$  \\[0.3mm]
          $\sigma_{8}$            & $0.961_{-0.064}^{+0.073}$  & $0.833\pm0.033$   &  $0.823\pm0.022$  \\[0.3mm]
            $h$                   & $0.810_{-0.062}^{+0.075}$  & $0.699\pm0.023$   &  $0.692\pm0.013$  \\[0.3mm]
        $t_{0}/{\rm Gyr}$         & $13.84\pm0.13$             & $13.76\pm0.15$    &  $13.71\pm0.13$  \\[0.3mm]
\hline
\end{tabular}
\label{tab:kwcdm}
\end{table*}

\begin{table*}  
\centering
  \caption{
    Marginalized 68\% constraints on the$\Lambda$CDM model extended by treating $\sum m_\nu$ as a free parameter,
obtained using different combinations of the data sets described in Section~\ref{sec:data}. 
}
    \begin{tabular}{@{}lccc@{}}
    \hline
 &\multirow{2}{*}{ePlanck+BOSS $\xi(s)$}  & \multirow{2}{*}{ePlanck+BOSS $\xi_{\Delta \mu}(s)$} & ePlanck + BOSS $\xi_{\Delta \mu}(s)$  \\
                                         &                       &                             &    +BAO+SN         \\  
\hline
$\sum m_\nu$             &  $ < 0.23\,{\rm eV} $ (95\% CL) & $< 0.24\,{\rm eV} $ (95\% CL)   & $ < 0.23\,{\rm eV}$ (95\% CL)\\ [0.3mm]
        $100\,\omega_{\rm b}$  & $2.221\pm0.023$   &  $2.226\pm0.024$  &  $2.228\pm0.024$    \\[0.3mm]
         $100\,\omega_{\rm b}$  & $2.221\pm0.023$   &  $2.226\pm0.024$  &  $2.228\pm0.024$    \\[0.3mm]
        $100\,\omega_{\rm c}$   & $11.76\pm0.17$    &  $11.67\pm0.18$   &  $11.64\pm0.17$    \\[0.3mm]
        $100\,\Theta$           & $104.16\pm0.056$  &  $104.17\pm057$      &  $104.17\pm0.056$    \\[0.3mm]
        $n_{\rm s}$             & $0.9631\pm0.0054$ &  $0.9651\pm0.0056$  &  $0.9656\pm0.0054$    \\[0.3mm]
      $\ln(10^{10}A_{\rm s})$   & $3.089\pm0.025$   &  $3.087\pm0.026$     &  $3.088\pm0.026$    \\[0.3mm]
\multicolumn{3}{l}{Derived parameters} &  \\[0.3mm]
         $f_\nu$                &  $< 0.017 $ (95\% CL)   & $< 0.019$ (95\% CL)     & $ < 0.017$ (95\% CL) \\[0.3mm]
         $100\Omega_{\rm DE}$   & $69.6\pm1.0$   &  $70.0\pm1.0$     &  $70.3\pm1.0$    \\[0.3mm]
         $100\Omega_{\rm m}$    & $30.4\pm1.0$   &  $30.0\pm1.0$     &  $29.7\pm1.0$    \\[0.3mm]
         $\sigma_{8}$           & $0.816\pm0.020$   &  $0.809\pm0.021$     &  $0.811\pm0.020$    \\[0.3mm]
         $h$      & $0.6808\pm0.0082$ &  $0.6840_{-0.0091}^{+0.0089}$  &  $0.6860_{-0.0086}^{+0.0085}$    \\[0.3mm]
       $t_{0}/{\rm Gyr}$        & $13.795\pm0.046$  &  $13.79\pm0.050$  &  $13.78_{-0.0047}^{+0.0048}$    \\
\hline
\end{tabular}
\label{tab:mnu}
\end{table*}

\begin{table*}  
\centering
  \caption{
The marginalized 68\% constraints on the cosmological parameters of the $\Lambda$CDM model extended by allowing
for simultaneous variations on $\sum m_\nu$ and $w_{\rm DE}$, obtained using different combinations of the datasets described in Section~\ref{sec:data}.
}
    \begin{tabular}{@{}lccc@{}}
    \hline
 &\multirow{2}{*}{ePlanck+BOSS $\xi(s)$}  & \multirow{2}{*}{ePlanck+BOSS $\xi_{\Delta \mu}(s)$} & ePlanck + BOSS $\xi_{\Delta \mu}(s)$  \\
                                         &                       &                             &    +BAO+SN         \\  
\hline
$\sum m_\nu$      &  $ < 0.49\,{\rm eV} $ (95\% CL) &  $ < 0.47\,{\rm eV} $ (95\% CL)  & $ < 0.33\,{\rm eV}$ (95\% CL)\\ [0.3mm]
            $w_{\rm DE}$             & $-1.49_{-0.30}^{+0.24}$  &  $-1.13\pm0.12$    &  $-1.046\pm0.064$    \\[0.3mm]
         $100\,\omega_{\rm b}$      & $2.203\pm0.024$          &  $2.219\pm0.025$   &  $2.223\pm0.025$    \\[0.3mm]
          $100\,\omega_{\rm c}$     & $12.03\pm0.21$           &  $11.75\pm0.20$    &  $11.70\pm0.20$    \\[0.3mm]
          $100\,\Theta$             & $104.12\pm0.058$         &  $104.15\pm0.060$  &  $104.16\pm0.059$    \\[0.3mm]
          $n_{\rm s}$               & $0.9565\pm0.0060$        &  $0.9627\pm0.0062$ &  $0.9642\pm0.0059$    \\[0.3mm]
        $\ln(10^{10}A_{\rm s})$     & $3.090\pm0.025$          &  $3.087\pm0.026$   &  $3.087\pm0.025$    \\[0.3mm]
\multicolumn{3}{l}{Derived parameters} &  \\[0.3mm]
           $f_\nu$                  &  $< 0.036 $ (95\% CL) & $< 0.035$ (95\% CL)   & $ < 0.025$ (95\% CL) \\[0.3mm]
           $100\Omega_{\rm DE}$     & $76.8_{-3.1}^{+4.0}$  &  $71.6\pm1.8$ &  $70.7\pm1.2$    \\[0.3mm]
           $100\Omega_{\rm m}$      & $23.2_{-4.0}^{+3.1}$  &  $28.4\pm1.8$ &  $29.3\pm1.2$    \\[0.3mm]
           $\sigma_{8}$             & $0.910\pm0.058$            &  $0.821\pm0.031$ &  $0.815\pm0.026$    \\[0.3mm]
             $h$                    & $0.796_{-0.058}^{+0.076}$  &  $0.707\pm0.025$ &  $0.693\pm0.014$    \\[0.3mm]
         $t_{0}/{\rm Gyr}$          & $13.71_{-0.69}^{+0.70}$    &  $13.79\pm0.064$ &  $13.78\pm0.054$    \\[0.3mm]
\hline
\end{tabular}
\label{tab:mnuw}
\end{table*}

\begin{table*}  
\centering
  \caption{
Marginalized 68\% constraints on the cosmological parameters of the $\Lambda$CDM model extended by including
$N_{\rm eff}$ as a free parameter, obtained using different combinations of the datasets described in Section~\ref{sec:data}.
}
    \begin{tabular}{@{}lccc@{}}
    \hline
 &\multirow{2}{*}{ePlanck+BOSS $\xi(s)$}  & \multirow{2}{*}{ePlanck+BOSS $\xi_{\Delta \mu}(s)$} & ePlanck + BOSS $\xi_{\Delta \mu}(s)$  \\
                                         &                       &                             &    +BAO+SN         \\  
\hline
          $N_{\rm eff}$          & $3.35\pm0.27$    &  $3.31\pm0.27$    &  $3.30\pm0.27$    \\[0.3mm]
        $100\,\omega_{\rm b}$   & $2.239\pm0.030$  &  $2.243\pm0.031$  &  $2.243\pm0.029$    \\[0.3mm]
        $100\,\omega_{\rm c}$   & $12.25\pm0.47$   &  $12.10\pm0.46$   &  $12.07\pm0.45$    \\[0.3mm]
        $100\,\Theta$           & $104.11\pm0.072$ &  $104.13\pm0.072$ &  $104.13\pm0.072$    \\[0.3mm]
        $n_{\rm s}$             & $0.972\pm0.010$  &  $0.973\pm0.011$  &  $0.973\pm0.010$    \\[0.3mm]
      $\ln(10^{10}A_{\rm s})$   & $3.108\pm0.030$  &  $3.100\pm0.029$  &  $3.101\pm0.029$    \\[0.3mm]
\multicolumn{3}{l}{Derived parameters} &  \\[0.3mm]
         $100\Omega_{\rm DE}$   & $70.3\pm1.0$     &  $70.8\pm1.1$     &  $70.9\pm1.0$    \\[0.3mm]
         $100\Omega_{\rm m}$    & $29.7\pm1.0$     &  $29.2\pm1.1$     &  $29.1\pm1.0$    \\[0.3mm]
         $\sigma_{8}$           & $0.839\pm0.019$  &  $0.831\pm0.018$  &  $0.831\pm0.018$    \\[0.3mm]
           $h$                  & $0.701\pm0.018$  &  $0.703\pm0.019$  &  $0.703\pm0.018$    \\[0.3mm]
       $t_{0}/{\rm Gyr}$        & $13.50\pm0.26$   &  $13.52\pm0.26$   &  $13.53\pm0.26$    \\
\hline
\end{tabular}
\label{tab:nnu}
\end{table*}

\begin{table*}  
\centering
  \caption{
    The marginalized 68\% constraints on the parameters of the $\Lambda$CDM model extended 
    by assuming $f(z_{\rm m})=\Omega_{\rm m}(z)^\gamma$ and treating $\gamma$ as a free parameter,
and when $\gamma$ and $w_{\rm DE}$ are varied simultaneously.
}
    \begin{tabular}{@{}lcccc@{}}
    \hline
  & \multirow{2}{*}{ePlanck+BOSS $\xi_{\Delta \mu}(s)$} & ePlanck + BOSS $\xi_{\Delta \mu}(s)$ &\multirow{2}{*}{ePlanck+BOSS $\xi_{\Delta \mu}(s)$} & ePlanck + BOSS $\xi_{\Delta \mu}(s)$ \\
   &                                                     &    +BAO+SN        &                                                     &    +BAO+SN       \\  
\hline
         $w_{\rm DE}$            &             --       &           --      & $-1.16\pm0.11$   &   $-1.055\pm0.058$    \\[0.3mm]
        $\gamma$                 &    $0.69\pm0.15$     &  $0.69\pm0.15$    & $0.89\pm0.22$    &   $0.76\pm0.17$      \\[0.3mm]
  $100\,\omega_{\rm b}$          &    $2.225\pm0.024$   &  $2.226\pm0.024$  & $2.213\pm0.025$  &  $2.220\pm0.025$      \\[0.3mm]
  $100\,\omega_{\rm c}$          &    $11.70\pm0.16$    &  $11.68\pm0.016$  & $11.91\pm0.21$   &   $11.79\pm0.20$      \\[0.3mm]
  $100\,\Theta$                  &    $104.17\pm0.055$  &  $104.17\pm0.055$ & $104.14\pm0.061$  &   $104.16\pm0.057$      \\[0.3mm]
  $n_{\rm s}$                    &    $0.9644\om0.0055$ &  $0.9648\pm0.0053$  & $0.9594\pm0.0062$ &   $0.9622\pm0.0058$      \\[0.3mm]
 $\ln(10^{10}A_{\rm s})$         &    $3.090\pm0.025$   &  $3.091\pm0.025$    & $3.088\pm0.025$   &   $3.088\pm0.025$      \\[0.3mm]
\multicolumn{3}{l}{Derived parameters} &  &\\[0.3mm]
   $100\Omega_{\rm DE}$          &    $0.7024_{-0.0095}^{+0.0092}$  &  $0.7038\pm0.0090$  & $0.728\pm0.021$   &   $0.711\pm0.012$      \\[0.3mm]
   $100\Omega_{\rm m}$           &    $0.2976_{-0.0092}^{+0.0095}$  &  $0.2962\pm0.0090$  & $0.272\pm0.021$   &   $0.289\pm0.012$      \\[0.3mm]
   $\sigma_{8}$                  &    $0.820\pm0.011$               &  $0.820\pm0.011$    & $0.870\pm0.037$   &   $0.838\pm0.023$      \\[0.3mm]
     $h$                         &    $0.6859_{-0.0074}^{+0.0073}$  &  $0.6869\pm0.0071$  & $0.724\pm0.030$   &   $0.699\pm0.014$      \\[0.3mm]
 $t_{0}/{\rm Gyr}$               &    $13.772\pm0.036$              &  $13.77\pm0.036$  & $13.718\pm0.052$   &   $13.753\pm0.040$      \\[0.3mm]
       $f(z=0.32)$               &    $0.6777_{-0.0084}^{+0.0087}$  &  $0.6764_{-0.0082}^{+0.0083}$  & $0.6800\pm0.0090$ &   $0.6789_{-0.0087}^{+0.0088}$      \\[0.3mm]
       $f(z=0.57)$               &    $0.7699_{-0.0071}^{+0.0073}$  &  $0.7688_{-0.0069}^{+0.0070}$  & $0.784\pm0.012$   &   $0.7754_{-0.0098}^{+0.0099}$    \\[0.3mm]
\hline
\end{tabular}
\label{tab:fg}
\end{table*}

\end{document}